\documentclass[
reprint,6,
superscriptaddress,amsmath,
amssymb,
aps,
pra,
floatfix,
]{revtex4-2}

\usepackage{graphicx}
\usepackage{dcolumn}
\usepackage{bm}
\usepackage{float}
\usepackage{physics}
\usepackage{xcolor, soul}

\usepackage[mathlines]{lineno}

\usepackage{wrapfig}
\usepackage{subcaption}
\begin{document}

\preprint{APS/123-QED}

\title{Multi-mode Perturbation Modelling for Cavity Polygon and Star Modes}

\author{Saeed Farajollahi}
\affiliation{%
Department of Electrical and Computer Engineering, University of Victoria, Victoria, British
Columbia V8P 5C2, Canada.
}
\author{Zhiwei Fang}
\affiliation{%
XXL—The Extreme Optoelectromechanics Laboratory, School of Physics and Electronic Science, East
China Normal University, Shanghai 200241, China.
}
\author{Jintian Lin}
\affiliation{%
State Key Laboratory of High Field Laser Physics and CAS Center for Excellence in Ultra-Intense
Laser Science, Shanghai Institute of Optics and Fine Mechanics (SIOM), Chinese Academy of Sciences
(CAS), Shanghai 201800, China.
}
\affiliation{%
Center of Materials Science and Optoelectronics Engineering, University of Chinese Academy of
Sciences, Beijing 100049, China.
}
\author{Shahin Honari}
\affiliation{%
Department of Electrical and Computer Engineering, University of Victoria, Victoria, British
Columbia V8P 5C2, Canada.
}
\author{Ya Cheng}
\email{ya.cheng@siom.ac.cn}
\affiliation{%
XXL—The Extreme Optoelectromechanics Laboratory, School of Physics and Electronic Science, East
China Normal University, Shanghai 200241, China.
}
\affiliation{%
State Key Laboratory of High Field Laser Physics and CAS Center for Excellence in Ultra-Intense
Laser Science, Shanghai Institute of Optics and Fine Mechanics (SIOM), Chinese Academy of Sciences
(CAS), Shanghai 201800, China.
}
\affiliation{%
Center of Materials Science and Optoelectronics Engineering, University of Chinese Academy of
Sciences, Beijing 100049, China.
}
\affiliation{%
State Key Laboratory of Precision Spectroscopy, East China Normal University, Shanghai 200062,
China
}

\affiliation{%
Collaborative Innovation Center of Extreme Optics, Shanxi University, Taiyuan 030006, China
}
\affiliation{%
Collaborative Innovation Center of Light Manipulations and Applications, Shandong Normal
University, Jinan 250358, China.
}
\author{Tao Lu}%
 \email{taolu@ece.uvic.ca}
\affiliation{%
Department of Electrical and Computer Engineering, University of Victoria, Victoria, British
Columbia V8P 5C2, Canada.
}

\begin{abstract}
Polygon and star modes enable unidirectional emission and single-frequency lasing in whispering gallery microcavities. To understand their properties and facilitate design, we have adopted both two-dimensional and three-dimensional full-wave perturbation methods to simulate these modes. Our simulation demonstrates that a tapered optical fiber can be used as a weak perturbation to coherently combine multiple whispering gallery modes into a polygon or star mode. Additionally, our simulation predicts an optical quality factor as high as $10^7$ for the polygon modes, which is in good agreement with the experiment results.
\end{abstract}

\maketitle

\section{\label{sec:introduction}Introduction}
Conventional optical whispering gallery microresonators are known for their high-quality factors (Q) and small mode volume, making them suitable candidates for a range of applications from sensing, frequency microcomb generation, quantum information to optomechanics~\cite{vahala2003optical,lu2011high,baaske2014single,kippenberg2011microresonator,kippenberg2011microresonator,kippenberg2008cavity,honari2021fabrication,yu2016cavity}. When the azimuthal symmetry is lifted through structural deformation, a whispering gallery microcavity may form chaotic modes that evolve to star or polygon modes under certain conditions.  In the past, intriguing phenomena such as unidirectional emission have been found in these cavities~\cite{redding2012local,fang2007control,unterhinninghofen2008goos,harayama2005theory,wiersig2008combining,rex2002fresnel}. However, due to large optical loss arising from deformation, the high quality factor needed for activating nonlinear optical effects does not materialize. Recently these modes were observed in lithium niobate (LN) microdisks for which the azimuthal symmetry is lifted through a weak perturbation from a tapered optical fiber~\cite{fang2020polygon,lin2022electro}. In contrast to deformed whispering gallery microresonators, these microdisks can have polygon and star modes whithout degrading Q. Hence, they can be used in novel applications such as second harmonic generation, optomechanical oscillation and frequency microcomb generation due to the high intracavity optical intensity.

For deformed cavities, the formation of these modes has been explained by describing internal ray dynamics using Poincar\'{e} surface of section~\cite{redding2012local,fang2007control,unterhinninghofen2008goos,harayama2005theory,wiersig2008combining,rex2002fresnel}. However, obtaining full vector, three dimensional (3-D) field profile, resonance wavelength and quality factor of these modes remains a challenge. One way to obtain these parameters is to perform a full wave first principal based 3-D simulations on the structure. However, such 3-D simulation of non-symmetric microcavities is computationally intensive. To accommodate this issue, two-dimensional (2-D) approximations such as effective index methods are often applied~\cite{lee2004quasiscarred,fang2020polygon,lin2022electro,wiersig2002boundary,zou2009accurately,zou2011quick}.  Although proven to be powerful for photonic structures with low refractive index contrast profile, such approximation yields large inaccuracy in the case of microcavity simulation due to the large refractive index difference between the cavity material and ambient air.

In the past, perturbation methods were used as a powerful tool for chaotic mode analysis in microcavities~\cite{teraoka2003perturbation,arnold2003shift,teraoka2006theory,foreman2013theory,swaim2011detection}. In those implementations, the field profile of a whispering gallery microcavity having structural deformation or in the presence of a perturbed object was represented as a linear superposition of ideal whispering gallery modes (WGMs) field profile whose resonance wavelengths are close to the probe laser wavelength~\cite{lai1990time,lee2008resonances} and non-zero superposition coefficients are only available at a discrete number of resonant wavelengths~\cite{lee2008resonances,tureci2005modes,korneev2016perturbation}. At certain wavelengths, different polygon and star modes can be formed.

In this paper, through first-order perturbation, we investigate polygon and star modes formation in fiber-perturbed cavities. In contrast to the previous formalisms~\cite{lee2008resonances,tureci2005modes,korneev2016perturbation}, we adopted non-integral azimuthal mode order of each WGM to improve the phase estimation accuracy. In addition to obtaining the field profile and resonance wavelength, we also formulated a robust algorithm for quality factor estimation~\cite{gorodetsky1999optical,lin2022electro}.

\section{Multi-Mode Perturbation Formalism}\label{sec:formalism}
\begin{figure}[H]
\centering
\includegraphics[width=\columnwidth]{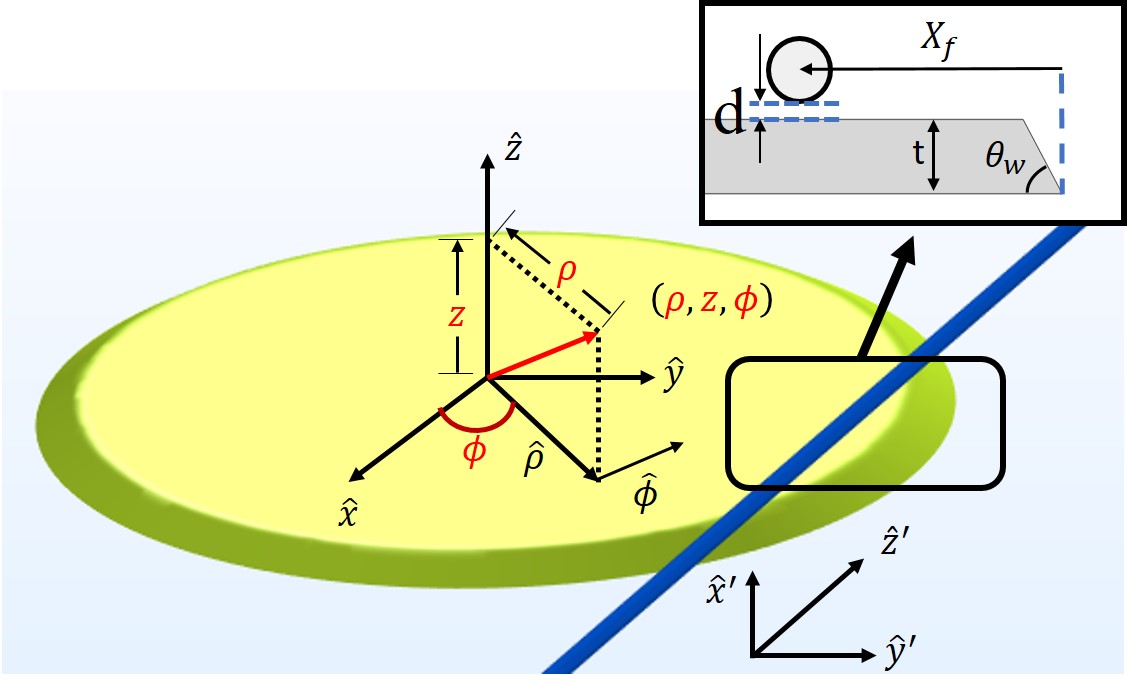}
\caption{\label{fig:str} Structure of a microdisk with radius $R$, thickness $t$ and wedge angle $\theta_w$ and a tapered fiber being placed on top of it. The vertical gap between the  fiber and the microdisk top surface is denoted by $d$.}
\end{figure}

The photonic device to be studied is shown in Fig.~\ref{fig:str}.  Here, a tapered fiber is placed near an azimuthal-symmetric z-cut LN microdisk with radius $R$, thickness $t$, and wedge angle $\theta_w$, and along $\hat x$, all are labeled in Fig.~\ref{fig:str}. The vertical gap between the fiber and the top surface of the disk is defined as $d$. $X_f$ is the horizontal distance from the center of fiber to the bottom edge of the microdisk at $\phi=\frac{\pi}{2}$ according to current settings. Here a negative $X_f$ represents the fiber being placed on top of the microdisk.

The multi-mode perturbation formalism starts with the Helmh\"{o}ltz equations that characterize the electric ${\vec E}(\rho,z,\phi$) and magnetic ${\vec H}(\rho,z,\phi$) field distributions of a non-magnetic photonic structure with relative permittivity profile $\epsilon_r(\rho,z,\phi$)
\begin{equation}
 \nabla^{2}\vec{E}(\rho,z,\phi)+\epsilon_{r}(\rho,z,\phi)k^2_0\vec{E}(\rho,z,\phi)={\vec 0}
 \end{equation}
 \begin{equation}
 \nabla^{2}\vec{H}(\rho,z,\phi)+\epsilon_{r}(\rho,z,\phi)k^2_0\vec{H}(\rho,z,\phi)={\vec 0}
   \label{eq1-2}
\end{equation}
Here, in a cylindrical coordinates ($\hat{\rho},\hat{z},\hat{\phi}$) specified in Fig.~\ref{fig:str}, the second order differential operator $\nabla^2\equiv\frac{1}{\rho}\frac{\partial}{\partial\rho}(\rho\frac{\partial}{\partial\rho})+\frac{1}{\rho^2}\frac{\partial^2}{\partial\phi^2}+\frac{\partial^2}{\partial z^2}$. $k_0=\frac{2\pi}{\lambda_0}$ is the free space wavenumber and $\lambda_0$ the vacuum wavelength of the probe laser. By definition, an ideal whispering gallery microcavity has an azimuthal independent relative permittivity profile $\epsilon_r(\rho,z,\phi)=\epsilon_r(\rho,z$). Through the separation of variables between $\phi$ and ($\rho,z$), one may find a discrete set of resonance wavelengths $\lambda_0\in\{\lambda_{\nu\mu}; \nu,\mu=1,2\ldots\}$ that lead to non-zero solutions $\vec{E}_{\nu\mu}(\rho,z,\phi$) and $\vec{H}_{\nu\mu}(\rho,z,\phi$) to the Helmh\"{o}ltz equations, which are called whispering gallery modes (WGMs) in the form of
\begin{equation}
   \vec{E}_{\nu\mu}(\rho,z,\phi)={\hat{e}}_{\nu\mu}(\rho,z)e^{-jm_{\nu\mu}\phi}
\end{equation}
   
\begin{equation}
 \vec{H}_{\nu\mu}(\rho,z,\phi)={\hat{h}}_{\nu\mu}(\rho,z)e^{-jm_{\nu\mu}\phi}
 \label{eq2-2}
\end{equation}
Here, the integer subscripts ($\nu,\mu$) represent the azimuthal and transverse mode order respectively. $m_{\nu\mu}$ is a complex number whose real part $Re\{m_{\nu\mu}\}=\nu$ is equal to the azimuthal mode order to satisfy the single value condition $\vec{E}_{\nu\mu}(\rho,\phi,z)\approx\vec{E}_{\nu\mu}(\rho,\phi+2\pi,z$) when neglecting the optical losses. The imaginary part $Im\{m_{\nu\mu}\}$ characterizes the loss and is related to the optical quality factor $Q$ according to $Q=\frac{Re\{m_{\nu\mu}\}}{2Im\{m_{\nu\mu}\}}$.  $\hat{e}_{\nu\mu}(\rho,z$) and $\hat{h}_{\nu\mu}(\rho,z$) are the electric and magnetic mode field distributions at transverse cross-section that satisfy the $\phi$-independent mode equations
\begin{equation}
\nabla^2_\perp\hat{e}_{\nu\mu}(\rho,z) + [\epsilon_r(\rho,z)k^2_{\nu\mu}-\frac{m^2_{\nu\mu}}{\rho^2}]\hat{e}_{\nu\mu}(\rho,z)={\vec 0}
\end{equation}
\begin{equation}
\nabla^2_\perp\hat{h}_{\nu\mu}(\rho,z) + [\epsilon_r(\rho,z)k^2_{\nu\mu}-\frac{m^2_{\nu\mu}}{\rho^2}]\hat{h}_{\nu\mu}(\rho,z)={\vec 0}
 \label{eq3-2}
\end{equation}
with $\nabla^2_\perp\equiv\frac{1}{\rho}\frac{\partial}{\partial\rho}(\rho\frac{\partial}{\partial\rho})+\frac{\partial^2}{\partial z^2}$ and $k_{\nu\mu}=\frac{2\pi}{\lambda_{\nu\mu}}$. When $Im\{m_{\nu\mu}\}$ is sufficiently small, WGMs are quasi-orthogonal and $\hat{e}_{\nu\mu}(\rho,z$) and $\hat{h}_{\nu\mu}(\rho,z$) are normalized such that
\begin{equation}
\pi\epsilon_0\iint\epsilon_r(\rho,z)\hat{e}_{\nu^\prime\mu^\prime}^*\cdot\hat{e}_{\nu\mu}\rho d{\rho}dz=\delta_{\nu\nu^\prime}\delta_{\mu\mu^\prime}.
 \label{eq4-2}
\end{equation}
with $\delta$ being the Kronecker delta. When a perturbation element such as a tapered fiber is placed in close proximity to the cavity, the relative permittivity becomes azimuthal dependent by a small amount  ($\Delta\epsilon_r(\rho,z,\phi)\ll\epsilon_r(\rho,z$)). A probe laser light at a wavelength $\lambda_l$ and wavenumber $k_l=\frac{2\pi}{\lambda_l}$ delivered to the cavity through the tapered fiber may distribute its energy to several WGMs coherently. Therefore, the field around the cavity can be expressed as
\begin{equation}
\vec{E}(\rho,z,\phi)=\sum_{\nu,\mu}a_{\nu\mu}{\hat e}_{\nu\mu}(\rho,z)e^{-jm^\prime_{\nu\mu}\phi}
\end{equation}
\begin{equation}
\vec{H}(\rho,z,\phi)=\sum_{\nu,\mu}a_{\nu\mu}{\hat h}_{\nu\mu}(\rho,z)e^{-jm^\prime_{\nu\mu}\phi}
\label{eq5-2}.
\end{equation}
with the summation over $N$ whispering gallery modes whose resonance wavelengths $\lambda_{\nu\mu}$ are close to $\lambda_l$. By assuming photons at both wavelengths, $\lambda_{\nu\mu}$ and $\lambda_l$ travel at the same optical path length and experience the same optical loss, we obtain~\cite{du2013full,du2014generalized}.
\begin{equation}
    Re\{m^\prime_{\nu\mu}\}=\frac{\lambda_{\nu\mu}}{\lambda_l}\nu
    \end{equation}
    \begin{equation}
    Im\{m^\prime_{\nu\mu}\}=Im\{m_{\nu\mu}\}
\label{eq6-2}.
\end{equation}
Unlike in~\cite{lee2008resonances} where azimuthal mode number for WGMs in Eq.~\eqref{eq5-2} does not change with wavelength, here we modify it to reduce the phase error. Substitute Eq.~\eqref{eq5-2} into  Eq.~\eqref{eq1-2}, we obtain
\begin{widetext}
\begin{equation}
\nabla^2\sum_{\nu\mu}a_{\nu\mu}\hat{e}_{\nu\mu}(\rho,z)e^{-jm^\prime_{\nu\mu}\phi}+[\epsilon_r(\rho,z)+\Delta\epsilon_r(\rho,z,\phi)]k^2_l\sum_{\nu\mu}a_{\nu\mu}\hat{e}_{\nu\mu}(\rho,z)e^{-jm^\prime_{\nu\mu}\phi}={\vec 0}
 \label{eq7-2}
\end{equation}
\end{widetext}

For simplicity, we relabel subscripts ($\nu,\mu$) with a single integer ($\gamma; \gamma\in1,2,\ldots,N$), replace all subscript pairs accordingly and define $\Delta k_\gamma$ and $\Delta m_\gamma$ according to
\begin{equation}
k_l=k_\gamma+\Delta k_\gamma
\end{equation}
\begin{equation}
m^\prime_\gamma= m_\gamma+ \Delta m_\gamma
\label{eq8-2}
\end{equation}
Following perturbation approximation, we keep the first order perturbation terms and neglect all higher order terms, obtain
\begin{widetext}
\begin{equation}
\sum_{\gamma}a_{\gamma}\{\nabla^2_\perp\hat{e}_{\gamma}(\rho,z)+[\epsilon_r(\rho,z)k^2_\gamma-\frac{m^2_\gamma}{\rho^2}]\hat{e}_{\gamma}(\rho,z)+[2\epsilon_r(\rho,z)k_\gamma\Delta k_\gamma+k^2_\gamma\Delta \epsilon_r(\rho,z,\phi)-\frac{2m_\gamma\Delta m_\gamma}{\rho^2}]\hat{e}_{\gamma}(\rho,z)\}e^{-jm^\prime_\gamma\phi}={\vec 0}
 \label{eq9-2}
\end{equation}
Since the first two terms on the left side of the equation vanish according to Eq.~\eqref{eq3-2}, we have
\begin{equation}
\sum_{\gamma}a_{\gamma}[2\epsilon_r(\rho,z)k_\gamma\Delta k_\gamma+k^2_\gamma\Delta \epsilon_r(\rho,z,\phi)-\frac{2m_\gamma\Delta m_\mu}{\rho^2}]\hat{e}_{\gamma}(\rho,z)e^{-jm^\prime_\gamma\phi}={\vec 0}
 \label{eq10-2}
\end{equation}
\end{widetext}
Finally, multiplying the remaining terms by $\{\hat{e}^*_{\gamma\prime}(\rho,z) e^{jm^\prime_{\gamma^\prime}\phi}; \gamma^\prime{\in}1,2,\ldots,N\}$ and integrating over full space, we get $N$ homogeneous linear equations with $N$ unknowns, which can be expressed in a matrix form
\begin{equation}
{\tilde \Lambda}(\lambda_l){\vec a}=\begin{bmatrix}
\Lambda_{11} & ... & \Lambda_{1N} \\ \vdots & \ddots & \vdots \\ \Lambda_{N1} & ... & \Lambda_{NN}
\end{bmatrix}
\begin{bmatrix}
a_1 \\ \vdots \\ a_N
\end{bmatrix}
= \begin{bmatrix} 0 \\ \vdots \\ 0
\end{bmatrix}
 \label{eq11-2}.
\end{equation}
Here, the matrix element is $\lambda_l$ dependent
\begin{widetext}
\begin{equation}
\Lambda_{\gamma^\prime\gamma}(\lambda_l)=\iiint\left[2\epsilon_r(\rho,z)k_\gamma\Delta k_\gamma+k^2_\gamma\Delta \epsilon_r(\rho,z,\phi)-\frac{2m_\gamma\Delta m_\gamma}{\rho^2}\right]
\hat{e}^*_{\gamma^\prime(\rho,z)}\cdot\hat{e}_\gamma(\rho,z) e^{-j(m^\prime_\gamma-m^\prime_{\gamma^\prime})\phi}{\rho}d{\rho}dzd\phi
 \label{eq12-2}.
\end{equation}
\end{widetext}
The coefficients vector ${\vec a}=[a_1,a_2,..,a_N]^T$ has nonzero values only for a discrete number of wavelength $\lambda_l$ where the corresponding determinant of the matrix is zero. The corresponding eigen-vector gives the amplitudes $a_\gamma$ in Eq.~\eqref{eq5-2} and determines the field profile of the mode. We normalize these amplitudes according to
\begin{equation}
\sum_{\gamma=1}^N|a_\gamma|^2=1 (J)
 \label{eq13-2}.
\end{equation}

\subsection{Quality Factor Estimation}

Since under current formalism, the resulting perturbed field is the superposition of unperturbed WGMs, therefore, the power coupled to the fiber can not be derived directly from the field pattern. Instead, the total quality factor ($Q_t$) has to be derived from a separate estimation of intrinsic Q ($Q_i$) and coupling Q ($Q_c$). Note

\begin{equation}
\frac{1}{Q_t}=\frac{1}{Q_i}+\frac{1}{Q_c}=\frac{P_{abs}+P_{rad}}{\omega\times{U_{cav}}}+\frac{P_{coupling}}{\omega\times{U_{cav}}}
 \label{eq2-3}
\end{equation}
where $P_{abs}$, $P_{rad}$, and $P_{coupling}$ are the power lost through absorption, radiation and coupling to the fiber. Here we ignore the scattering-induced loss, which can be easily estimated through Mie or Rayleigh scattering theory. $P_{abs}$, $P_{rad}$ and the resulting $Q_i$ can be estimated from the field distributions (see details in Supplementary Information, Section~S.1). Alternatively, sine the polygon mode is a linear combination of unperturbed whispering gallery modes (WGMs) in the absence of the coupling loss, $Q_i$ can also be derived from the complex resonance wavelength $\lambda_l$ in Eq.~\eqref{eq11-2}
\begin{equation}
Q_i=\frac{Re\{{\lambda_l}\}}{2Im\{{\lambda_l}\}}
 \label{eq3-1-3}.
\end{equation}

To estimate the coupling loss, the coupled mode theory (CMT) needs to be used~\cite{rowland1993evanescent}. Assuming a Cartesian coordinate system ($x^\prime,y^\prime,z^\prime$) for the fiber with ${\hat z}^\prime$ the direction of propagation as shown in Fig.~\ref{fig:str}, the fiber field can be written as

\begin{equation}
\vec{E}_f(x^\prime,y^\prime,z^\prime)  = b(z^\prime)\hat{e}_f(x^\prime,y^\prime)
\end{equation}
\begin{equation}
\vec{H}_f(x^\prime,y^\prime,z^\prime)  = b(z^\prime)\hat{h}_f(x^\prime,y^\prime)
 \label{eq5-3}
\end{equation}
Here, the fiber electric and magnetic mode field distributions $\{{\hat e}_f(x^\prime,y^\prime),{\hat h}_f(x^\prime,y^\prime)\}$ are normalized such that the power propagating along $z^\prime$ direction is $|b(z^\prime)|^2$. As $b(-\infty)=0$, $b(+\infty$) only contains the field coupled to the fiber output from the cavity~\cite{snyder2012optical,gorodetsky1999optical}, we have
\begin{widetext}
\begin{equation}
|b(z^\prime)|^2 = \left|\frac{k_0}{4\eta_0}\int^{z^\prime}_{-\infty}dz^{\prime\prime}\iint\vec{E}(x^{\prime\prime},y^{\prime\prime},z^{\prime\prime})\cdot[\Delta{\tilde \epsilon_{f,r}}\hat{e}_{f}^*(x^{\prime\prime},y^{\prime\prime})]e^{j\beta_fz^{\prime\prime}}dA\right|^2
 \label{eq7-3}
\end{equation}
\end{widetext}

Where $\beta_f$ is the propagation constant for the fiber mode. Coupling loss then can be derived from $P_{coupling} = |b(+\infty)|^2$.  A detailed discussion on coupling $Q$ calculation can be found in Supplementary Information, Section~S.2.
\begin{figure*}[tbh!]
    \centering
    \includegraphics[width=\textwidth]{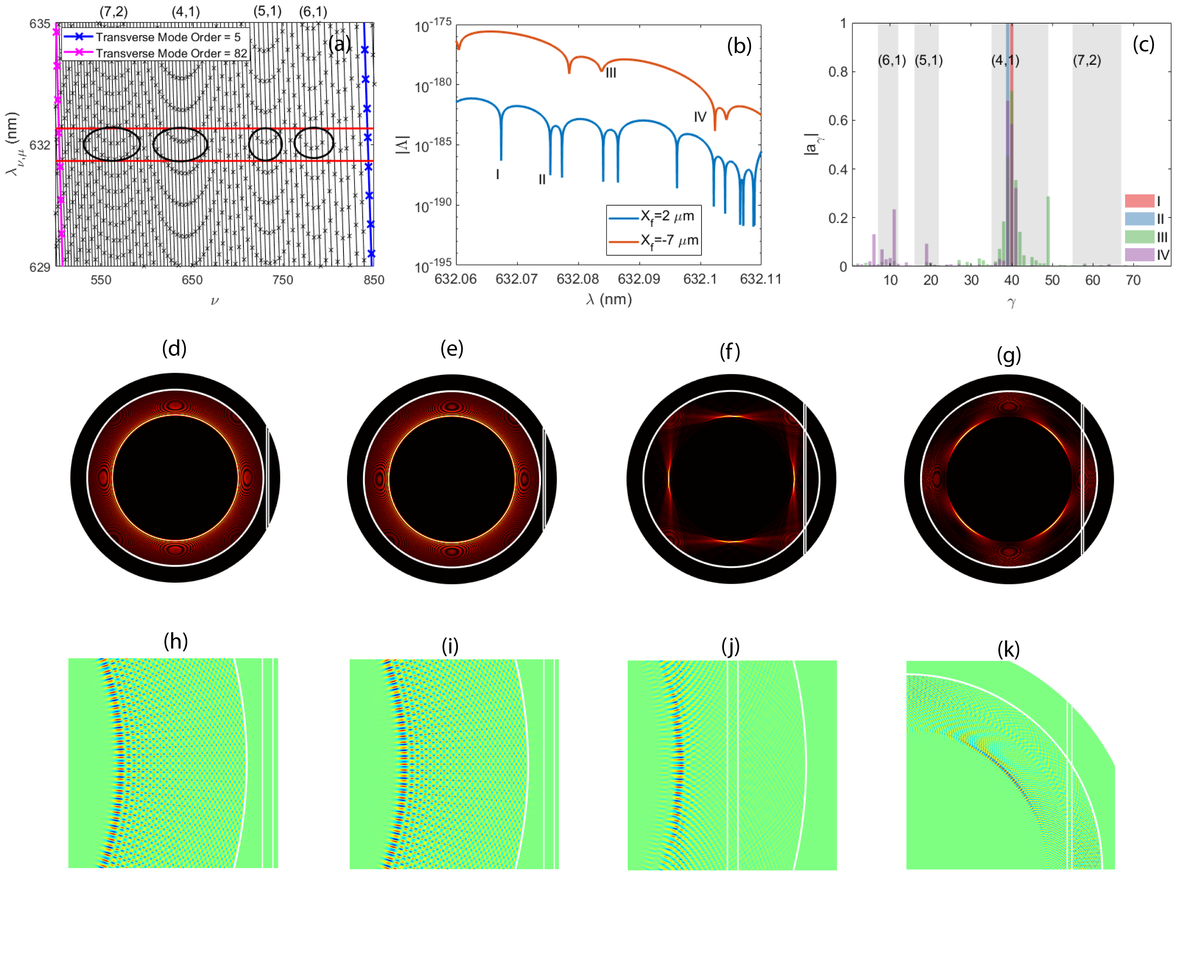}
    \caption{\label{Fig_35} (a) The resonance wavelengths of TE WGMs at different azimuth mode orders $\nu$. Black lines connect the same transverse mode order $\mu$. Transverse mode orders of 5 and 82 are shown with pink and blue lines. Sub-sets of WGMs with ($p,q$) pairs of ($7,2$), ($4,1$), ($5,1$) and ($6,1$) are shown in the figure with black circles. (b) The determinant $|det({\tilde \Lambda})|$ in the range $[632.05,632.11]$~nm. For the first case (orange line), $X_f = -7~\mu m$ and for the second case (blue line), $X_f = 2~\mu m$. (c) The amplitudes $|a_\gamma|$ for the modes I to IV in labelled in subplot (b). I-(g) The intensity profiles of the modes. (h)-(k) The zoom-in plots of electric field profiles $E_z$ for each mode. }
\end{figure*}

\subsection{2-D Simplification} Under 2-D approximation by using the effective index method to reduce the three-dimensional photonic structure to a two-dimensional one with the refractive index profile invariant along ${\hat z}$ direction ($\frac{\partial}{\partial z}\equiv 0$), WGMs decouple to ${\hat z}$-independent TE and TM modes with nonzero field profiles $\{e_{\gamma,z}(\rho),h_{\gamma,\rho}(\rho),h_{\gamma,\phi}(\rho)\}$ and $\{h_{\gamma,z}(\rho),e_{\gamma,\rho}(\rho),e_{\gamma,\phi}(\rho)\}$ respectively. The polygon and star modes can still be obtained following the formalism above except that the matrix elements in Eq.~\eqref{eq12-2} should be obtained from
\begin{widetext}
\begin{equation}
\Lambda_{\gamma^\prime\gamma}=\int_{\phi=0}^{2\pi}\int_{\rho=0}^{\infty}[2\epsilon_r(\rho,z)k_\gamma\Delta k_\gamma+k^2_\gamma\Delta \epsilon_r(\rho,\phi)-\frac{2m_\gamma\Delta m_\gamma}{\rho^2}]\hat{e}^*_{\gamma^\prime}(\rho)\cdot\hat{e}_\gamma(\rho) e^{-j(m^\prime_\gamma-m^\prime_{\gamma^\prime})\phi}\rho d\rho d\phi
 \label{eq14-2}
\end{equation}
\end{widetext}
In general, the 2-D WGM field profiles can be obtained numerically according to
\begin{equation}
\frac{1}{\rho}\frac{\partial}{\partial\rho}(\rho\frac{\partial}{\partial\rho})\hat{e}_{\nu\mu}(\rho) + [\epsilon_r(\rho)k^2_{\nu\mu}-\frac{m^2_{\nu\mu}}{\rho^2}]\hat{e}_{\nu\mu}(\rho)={\vec 0}
\end{equation}
\begin{equation}
\frac{1}{\rho}\frac{\partial}{\partial\rho}(\rho\frac{\partial}{\partial\rho})\hat{h}_{\nu\mu}(\rho) + [\epsilon_r(\rho)k^2_{\nu\mu}-\frac{m^2_{\nu\mu}}{\rho^2}]\hat{h}_{\nu\mu}(\rho)={\vec 0}
 \label{eqadd2}
\end{equation}
 In the special case where the wedge angle $\theta_w=90^\circ$, the analytic solution is obtainable~\cite{lee2008resonances} as shown in Supplementary Information Section~S.3. In this case, Eq.~\eqref{eq14-2} can be solved through the analytical mode solution efficiently.

\section{Results and Discussion}

\subsection{2-D Perturbation Results}
\begin{figure}[tbh!]
        \includegraphics[width=0.45\textwidth]{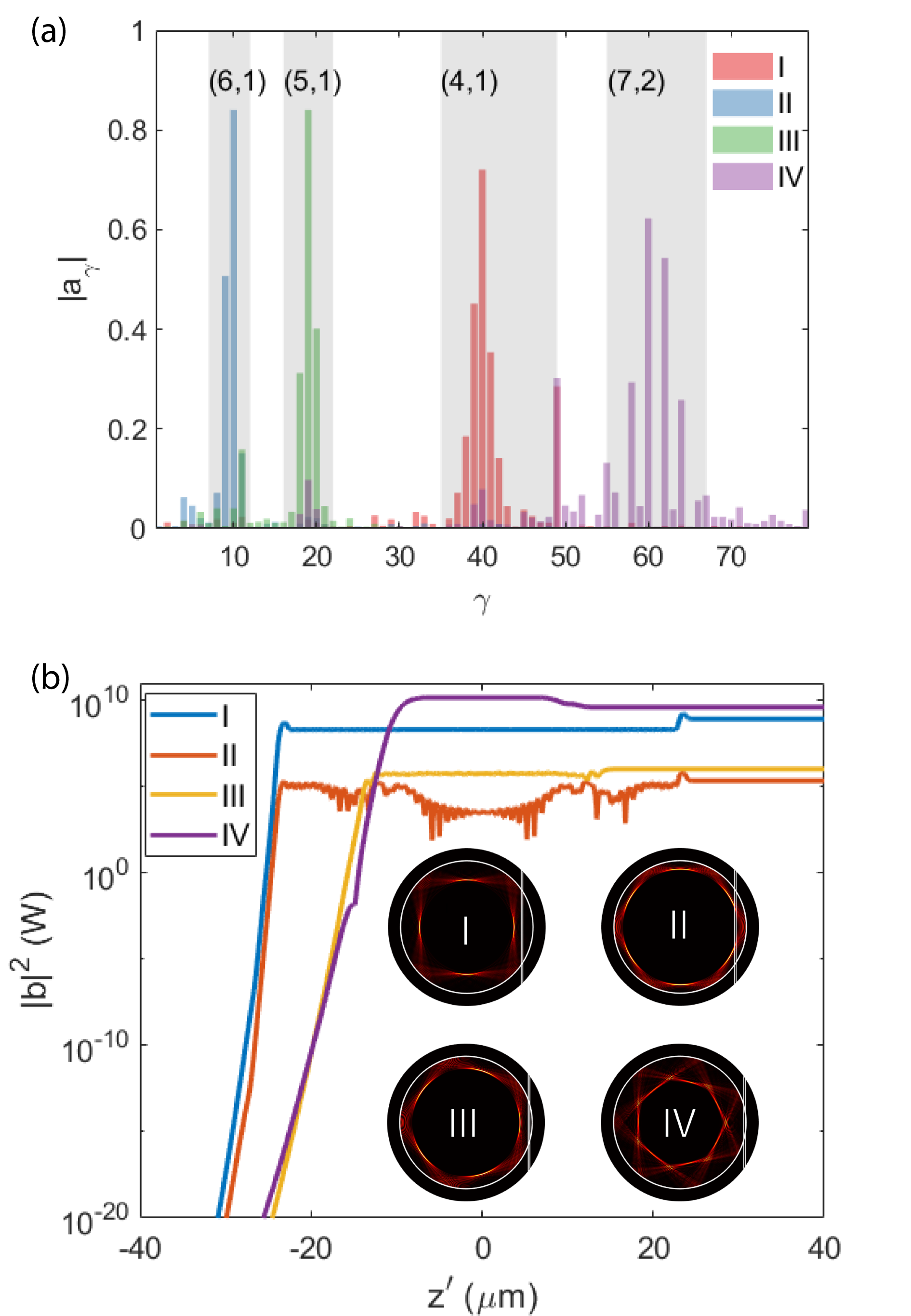}
        \caption{\label{fig:2d4modes}(a) $|a_\gamma|$ of the square (I), hexagon (II), pentagon (III) and heptagram (IV) modes when $X_f=-7~{\rm \mu}$m for I and II, while $X_f=-2~{\rm \mu}$m for III and $X_f=-1~{\rm \mu}$m for IV. (b) Optical power ($|b(z^\prime|^2$) propagating along the tapered fiber in the absence of light from its input end. Insets are the top-view intensity profiles of these modes. The coupling Q of these modes are $3.6\times10^6$, $1.4\times10^{10}$, $2.9\times10^9$ and $7.4\times10^5$ respectively.}
    \end{figure}

We first model polygon modes under the 2-D approximation and assume the wedge angle $\theta_w=90^\circ$ so that the analytical solution of WGMs can be adopted.  The disk geometry is identical to the actual disk demonstrated in~\cite{fang2020polygon} with $R=42~{\mu}m$ and at wavelengths around $632~nm$. Fig.~\ref{Fig_35}(a) shows the resonance wavelengths of unperturbed TE WGMs as a function of the azimuthal mode order $\nu$. Straight lines connect the modes with the same transverse mode order $\mu$. In particular, $\mu=5$ and $\mu=82$ are shown as blue and pink lines in the plot and the black lines between them correspond to modes with $\mu$ in between. The WGMs with azimuthal and transverse mode orders of $\nu(n)=\nu_0+n\times p$ and $\mu(n)=\mu_0-n\times q$ are nearly degenerate for ($p,q$) pairs of ($7,2$), ($4,1$), ($5,1$) and ($6,1$) at the sub-sets shown within the black ellipse in Fig.~\ref{Fig_35}(a). Here, $n$ is an integer and $\nu_0$ and $\mu_0$ are the azimuthal and transverse mode order of the first member in the subset. The near degeneracy of certain sub-sets of WGMs would lead to solutions for which the only non-negligible coefficients in Eq.~\eqref{eq5-2} belong to this sub-set. The shape of the resulting polygon or star mode then can be predicted from ($p,q$) pair. $p$ shows the number of bouncing points along the boundary and $q$ shows the number of windings until we return to the initial point~\cite{lee2008resonances}. Therefore, sub-sets with $q=1$ correspond to polygon modes and sub-sets with $q>1$ correspond to star modes.

\begin{figure*}[htb!]
\centering
\includegraphics[width=0.9\textwidth]{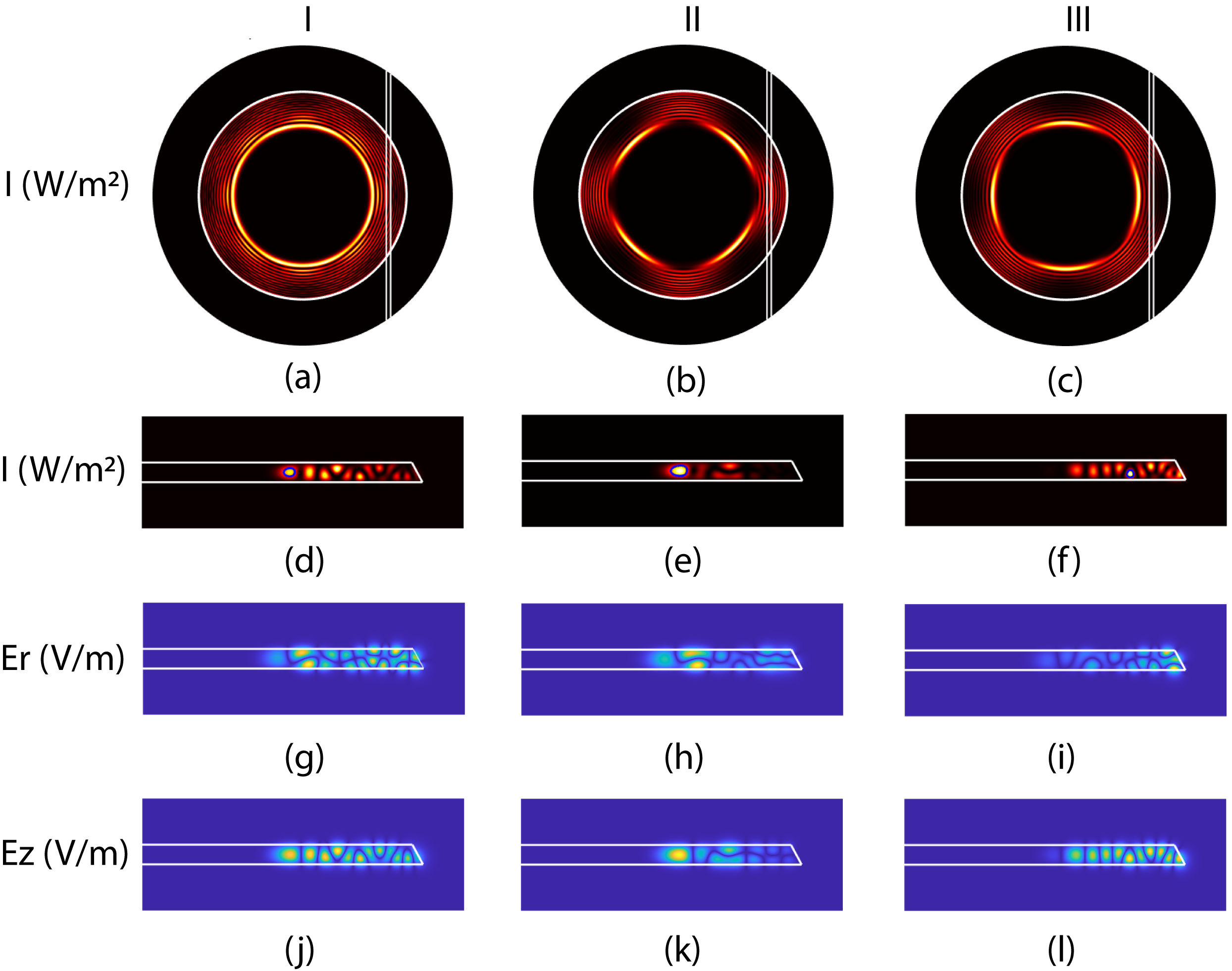}
\caption{\label{fig_pert3D2} The top view intensity profile at $z=0$ of (a) star mode  (b) square mode at $45^\circ$ and (c) horizontal square mode. Their resonant wavelength and quality factor ($\lambda_l,Q$) are ($972.4097~nm,7.1\times10^7$), ($972.4165~nm,4.6\times10^6$) and ($972.4314~nm,2.1089\times10^6$) respectively. The intensity (I), electrical field along radial ($E_r$) and vertical ($E_z$) at $\phi=45^\circ$ are displayed in (d)-(f), (g)-(i) and (j)-(l).}
\end{figure*}

In general, to find the mode of a particular polygon shape, it is sufficient to select the WGMs of the corresponding ($p,q$) pairs within the black circle in Fig.~\ref{Fig_35}(a). Here, for demonstration purposes, we select all the 79 WGMs whose transverse mode orders are between 5 and 82, and resonance wavelengths within one mean free spectral range (FSR, $0.8~nm$) of the center wavelength ($632\pm{0.4}$~nm, shown between the horizontal red lines in Fig.~\ref{Fig_35}(a)).  These modes are inserted into Eq.~\eqref{eq5-2} for two different cases.  In the first case, fiber is placed on top of the microdisk with $X_f=-7~{\rm \mu}$m and in the second case, fiber is placed far away from the microdisk with $X_f=+2~{\rm \mu}$m. Fig.~\ref{Fig_35}(b) shows the determinant $|{\tilde \Lambda}|$ vs. $\lambda_l$ for both cases.  As shown, a set of dips can be found in both the first case (orange solid line) and the second case (blue solid line). Each dip represents a resonant mode of the perturbed cavity since $|{\tilde \Lambda}|=0$ leads to non-zero solutions to ${\vec a}$ that are the coefficients of the contributing WGMs coherently forming the resonance mode. Here, we selected two modes in each case and labeled them by I-IV on the plot for further investigation.

The WGMs amplitudes $|a_\gamma|$ for all the four labeled modes are shown in  Fig.~\ref{Fig_35}(c).  In the second case where $X_f=+2~\mu m$, the coupling quality factor, $Q_c$ for modes I and II are $9.3\times10^{19}$ and $2\times10^{20}$, indicating a weak interaction between the fiber and disk. Consequently, only one non-zero $a_\gamma$ is present for mode I (orange bar) and mode II (blue bar), leading to pure WGM modes as shown from the intensity distributions in Fig.~\ref{Fig_35}(d)-(e) since the perturbation is too weak to excite multiple WGMs. For $X_f=-7~\mu m$ where the perturbation is relatively strong, $Q_c$ of mode III and IV are $3.6\times10^6$ and $7.2\times10^7$ respectively. With such strong fiber-to-disk interactions, there are multiple non-zero $a_\gamma$ values that coherently combine the WGMs into a polygon as shown in Fig.~\ref{Fig_35}(f)-(g). The intensity of these modes is shown in Fig.~\ref{Fig_35}(d)-(g) with the corresponding zoom-in images of electric field distributions in Fig.~\ref{Fig_35}(h)-(k). For the square modes III and IV, the non-zero amplitudes around $\nu = 40$ correspond to the sub-set with ($p,q)=(4,1$).   The total quality factor, $Q_t$ for modes I-IV are $1.1\times10^8$, $1.1\times10^8$, $3.4\times10^6$ and $4.6\times10^7$ respectively.

To obtain polygon and star modes of different shapes, we placed the fiber at different $X_f$ with fiber in direct contact with the disk top ($d=0$). As shown in Fig.~\ref{fig:2d4modes}(a), when $X_f$ is at $-7~{\rm \mu}$m, two modes were observed at resonance wavelengths $\lambda_l$ of $632.0838$~nm and $632.0107$~nm and labeled as I and II. The significant values of ${\vec a}$ fall into ($p,q$)$=$($4,1$) and ($6,1$) respectively, suggesting the creation of square and hexagon modes.   The top view intensity profiles showing as the insets in Fig.~\ref{fig:2d4modes}(b) confirmed that these two modes are the shapes suggested. We further placed the fiber at $X_f=-2~{\rm \mu}$m to find a pentagon mode (III) at $\lambda_l=632.067$~nm with ($p,q)=(5,1$). Finally a heptagram mode (IV) was located at $X_f=-1~{\rm \mu}$m with ($p,q)=(7,2$) and $\lambda_l=632.0627$~nm, both of which are also confirmed by the insets in Fig.~\ref{fig:2d4modes}(b). The main plot of Fig.~\ref{fig:2d4modes}(b) further shows the optical power coupled to the fiber ($|b(z^\prime)|^2$) for these modes. The coupling Q calculated for the modes I to IV based on $b(+\infty$) is $3.6\times10^6$, $1.4\times10^{10}$, $2.9\times10^9$ and $7.4\times10^5$ respectively. It is worth mentioning that for mode II and III, even though the fiber is placed in contact with the disk surface and the perturbation is strong, the overlap between the mode intensity to the fiber remains small. Consequently, the coupling loss is low, making a high overall quality factor possible. In contrast, a normal WGM would have low coupling Q due to over-coupling when the fiber is at the same location. Our simulation results further confirm the experiment observation in~\cite{fang2020polygon,lin2022electro} that a polygon mode may have a high $Q_t$ even when the fiber is placed on disk top.

\subsection{3-D Perturbation Results}

 In this subsection, we apply the 3-D full vector perturbation to polygon mode analysis. Here, we adopt a disk similar to the one that has been experimentally demonstrated in~\cite{lin2022electro} with $R=14.53~{\rm \mu}$m, $t=700$~nm, a wedge angle $\theta_w=61.6^\circ$, and at wavelengths around $970$~nm. The fiber has a diameter of $600~nm$ centered at $X_f=-2.53~\mu$m. The vertical gap between the fiber and the microdisk is $d=150~nm$. In total, we illustrated three modes by sweeping the wavelength from $972.39$~nm to $972.53$~nm  and labeled them as I, II and III in Fig.~\ref{fig_pert3D2}. The top-view intensity profiles of these modes at $z=0$ cross-section in Fig.~\ref{fig_pert3D2}(a)-(c) show that mode I is a weakly formed star mode while mode II and III are square modes, one oriented $45^\circ$ from ${\hat x}$ while the other is horizontal. The side-view at $\phi=45^\circ$ of intensity ($I$), electrical fields along radial ($E_r$) and vertical ($E_z$) directions are shown on Fig.~\ref{fig_pert3D2}(d)-(f), (g)-(i) and (j)-(l) respectively. The total Quality factor ($Q_t$) for the modes (I)-(III) are $7.1\times10^7$, $4.6\times10^6$ and $2.1\times10^6$ respectively. A video showing the 3-D field distribution is available in the Supplementary Material.

\subsection{Comparsion between 2-D and 3-D approaches}
\begin{figure}[tbh!]
\centering
\includegraphics[width=0.45\textwidth]{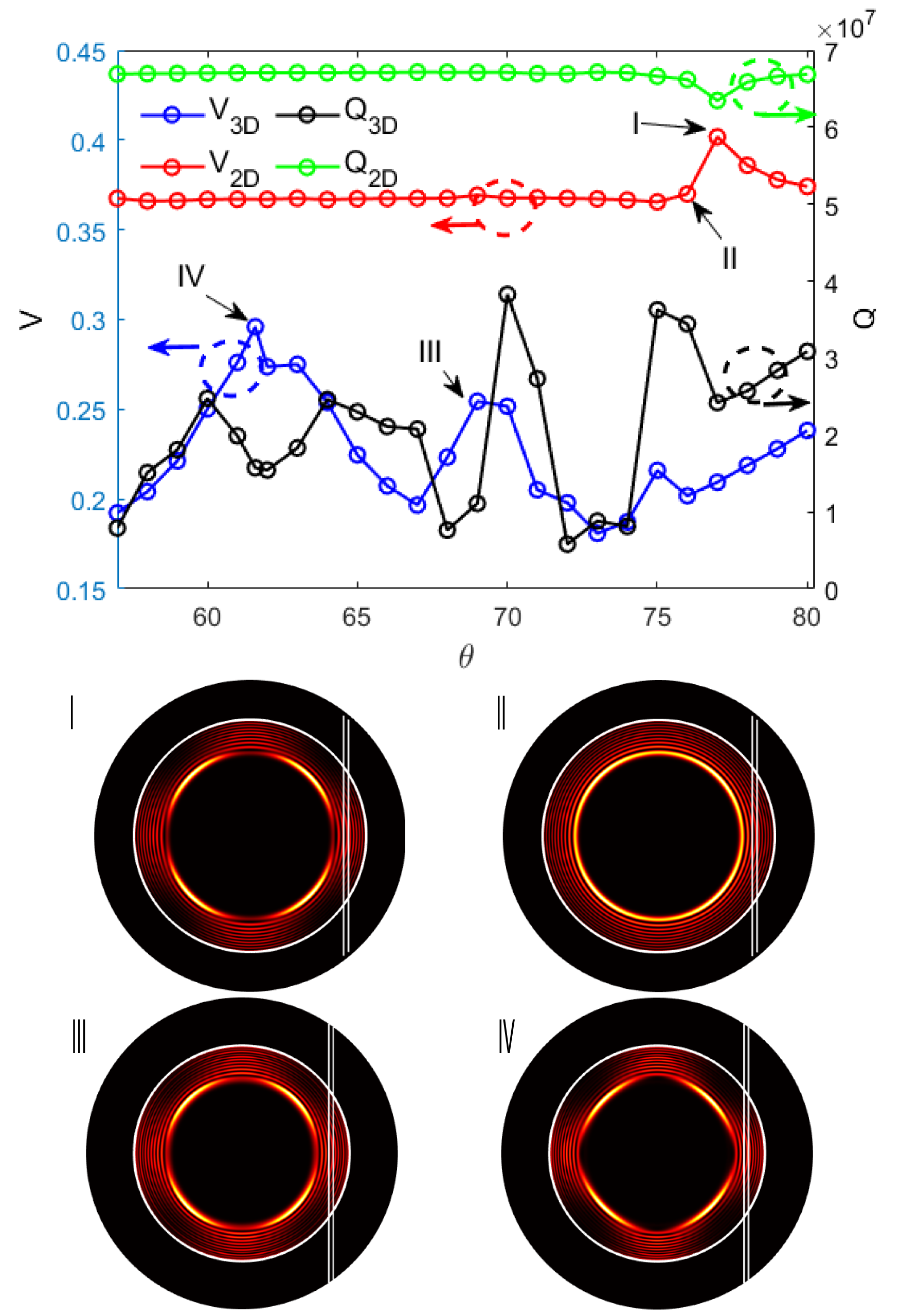}
\caption{\label{fig:VW} (a) $V$ as a function of wedge angle calculated for the same structure using 2D and 3D methods. (b) The distance between resonance wavelengths of ($p,q)=(4,1$) set in 2D. (c) $\lambda_s^{(4,1)}-\lambda_2^{(4,1)}$ in the 3D case.}
\end{figure}
\begin{figure}[tbh!]
\centering
    \begin{subfigure}[b]{0.45\textwidth}
        \includegraphics[width=0.8\textwidth]{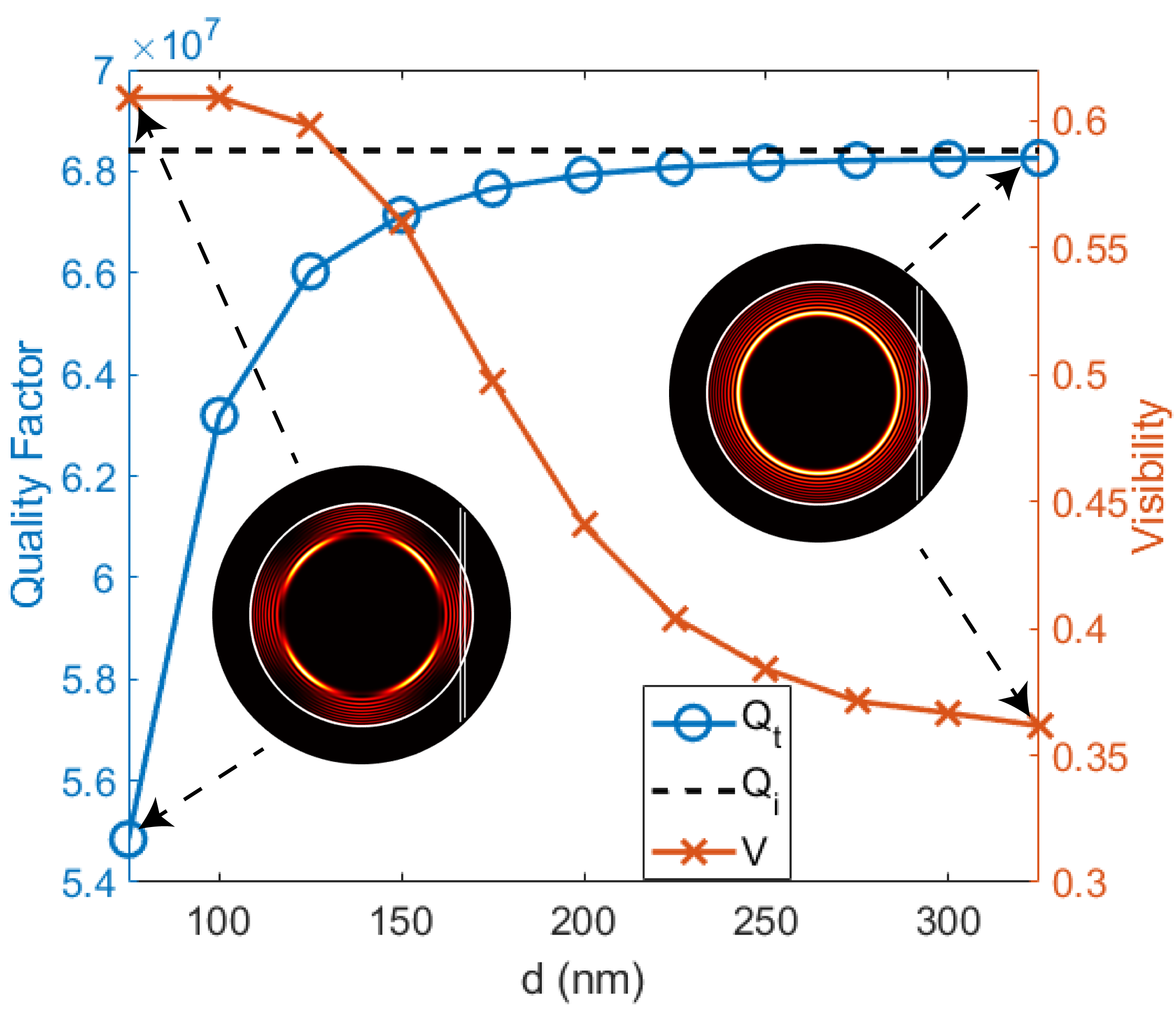}
        \caption{2-D perturbation}
    \end{subfigure}
   \begin{subfigure}[b]{0.45\textwidth}
        \includegraphics[width=0.8\textwidth]{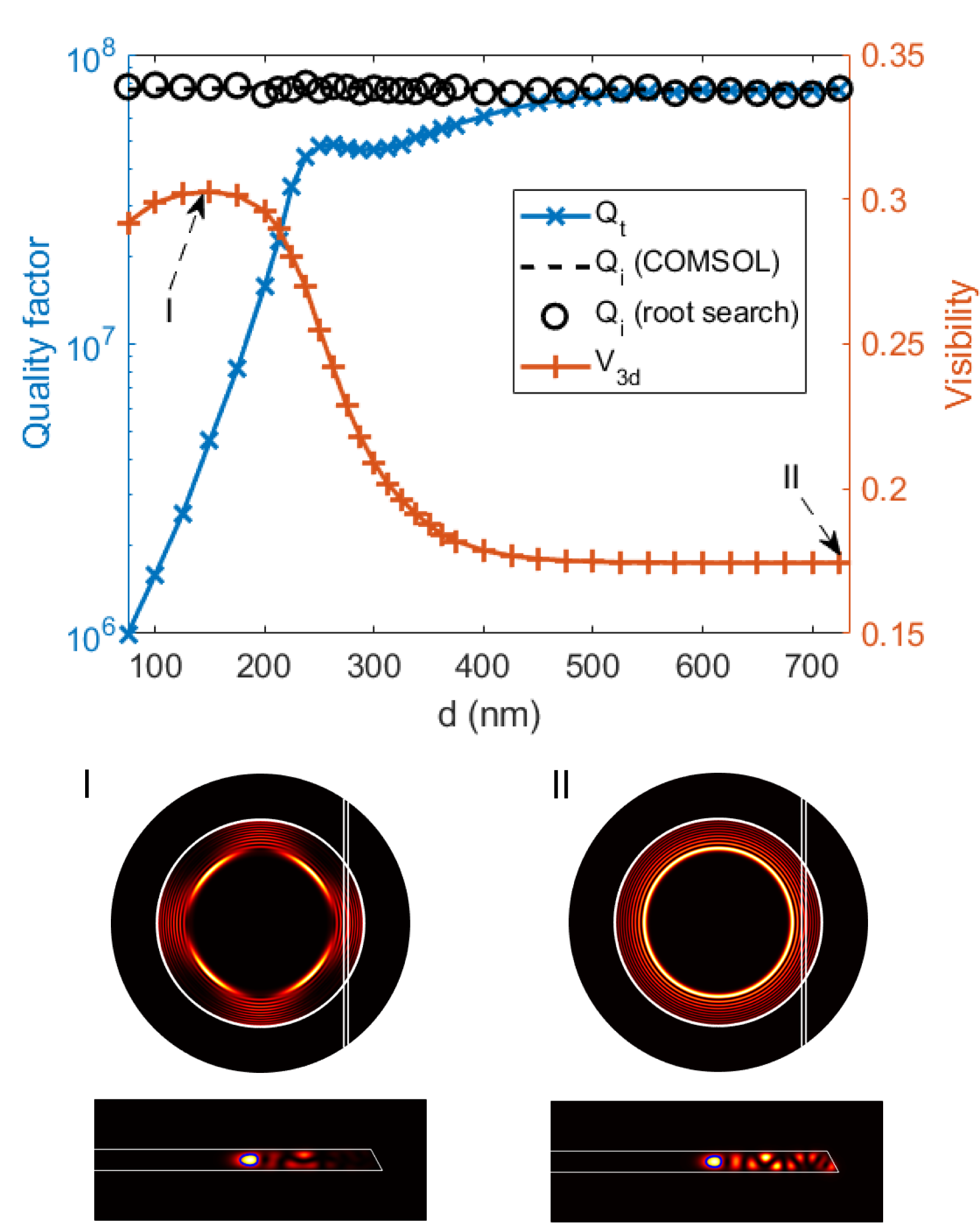}
        \caption{3-D perturbation.}
    \end{subfigure}
\caption{\label{Fig3_2d_4shape}  (a) 2-D perturbation simulation of quality factor and visibility evolution of the square polygon mode to WGM of a disk with wedge angle of $77^\circ$. The black dashed line shows the intrinsic quality factor of the corresponding WGM when $d$ is sufficiently large. The left inset is the intensity distribution of the square mode with the highest visibility of $0.6$ and $Q_t=5.5{\times}10^7$ at $d=75$~nm while the right inset displays the intensity distribution of the WGM at $d=325$~nm with $V_{2d}=0.36$ and $Q_t=6.8{\times}10^7$. (b) 3-D perturbation simulation of a disk similar to (a) except for a wedge angle of $61.6^\circ$. The dashed black line shows the intrinsic quality factor calculated from Eq.~\eqref{eq3-3} and the circles show the calculated intrinsic quality factor from Eq.~\eqref{eq3-1-3}. The top-view and side view intensity profiles of modes are also shown at $d=150$~nm (I) and $d=725$~nm (II). }
\end{figure}
To compare with 2-D perturbation method,  we further investigate the polygon mode evolution by varying the slant wedge angle $\theta_w$ and setting $d=200~nm$.

In 2-D perturbation, the analytic solution of 2-D WGMs is no longer available due to the appearance of the slant wedge angle and we adopted the numeric solution through COMSOL instead to obtain the WGMs (see Supplementary Information Section~S.4).  To evaluate the quality of the square mode, we defined a 2-D visibility $V_{2d}$ at $\phi=45^\circ$ according to
\begin{equation}
V_{2d}=\frac{\int_{\rho_{h1}}^{\rho_{h2}}I(\rho,\phi=45^\circ)d\rho}{\int_{-\infty}^{\infty}I(\rho,\phi=45^\circ)d\rho}
 \label{eq9-3}.
\end{equation}
Here $\rho_{h1}$ and $\rho_{h2}$ are the locations of the full width at half maximum (FWHM) of the mode intensity $I(\rho,\phi=45^\circ$) at $\phi=45^\circ$. In this simulation, we sweep the wedge angle from $57^\circ$ to $80^\circ$ to obtain a resonant mode at each angle. Using a Windows-10 computer with single Intel i7-7700K CPU operating at 4.20GHz and 64 GB memory, it takes around $850$ seconds to obtain each resonant mode. Shown as the red curve in Fig.~\ref{fig:VW}, the resonance modes are WGM in nature when we start to increase $\theta_w$ till $76^\circ$, as evident from the top-view intensity profile below the main plot (II). When the wedge angle reaches $77^\circ$, a sharp rise of visibility to $0.4$ suggests the formation of a square mode, which is confirmed by the intensity profile (I).

In 3-D, similar to 2-D case, we defined a 3-D visibility $V_{3d}$ of the square modes as

\begin{equation}
V_{3d}=\frac{\int\int_{S_h}I(\rho,z,\phi=45^\circ)d\rho dz}{\int_{-\infty}^{\infty}\int_{-\infty}^{\infty}I(\rho,z,\phi=45^\circ)d\rho dz}
 \label{eq1-3}.
\end{equation}

Here, $I(\rho,z,\phi$) is the intensity profile and $S_h$ is the area where the intensity is above the half maximum intensity. Similar to the 2-D simulation, we performed a wedge angle sweep to compute the resonant modes accordingly. Using the same computer, each mode computation takes around $5,700$ seconds. In this simulation, we observed two clear polygon modes with high visibility at $\theta=61.6^\circ$ and $\theta=67^\circ$. The higher visibility at $\theta=61.6^\circ$ is due to the fact that the resonance wavelengths of the contributing WGMs are closer (see Supplementary Information Section S.5). Note that the large departure of the wedge angle between the 2-D and 3-D simulation suggests that although 2-D is a fast algorithm with reasonable accuracy in generating field patterns, it is less accurate in the presents of highly sensitive parameters such as the wedge angle. A video showing the polygon mode evolution vs. the wedge angle is available in the Supplementary Material.

We further study the evolution of the polygon modes by varying the fiber-to-disk distance $d$, with the wedge angle in 2-D and 3-D simulations set at each of their optimum value ($77^\circ$ for 2-D and $61.6^\circ$ for 3-D). Further, we optimize the $X_f=-1.33~{\rm \mu}$m in 2-D simulation so that the visibility reaches the highest.  Fig.~\ref{Fig3_2d_4shape}(a) shows $Q_t$ (blue curve with circle markers to the left axis) and $V_{2d}$ (yellow curve with cross markers to the right axis) vs. $d$. The right inset of the intensity distribution at $d=325$~nm indicates that when gap between the fiber and the disk top is larger than half wavelength, the perturbation is negligible and only WGM can be formed with a poor square mode visibility of $0.36$ and a total quality factor $Q_t=6.8{\times}10^7$ that is dominated by $Q_i$ shown as the black dashed line. When the fiber moves closer to the disk top with decreasing $d$, the resonance mode gradually evolves to the square mode with increasing visibility till reaches the best square shape at $=75$~nm (left inset) at a resonance wavelength of $971.4453$~nm and with visibility of $0.6$ as a result of increased perturbation. On the other hand, due to the increased power coupling to the fiber, the $Q_t$ drops to $5.5{\times}10^7$.

In 3-D simulation, according to the visibility curve (orange line with plus markers), the sharpest square mode occurs at $d=150~nm$, almost twice as far as the 2-D case. It is further confirmed that although 2-D perturbation provides fast modeling of the cavity, it can not accurately predict the exact wedge angle and fiber position as these parameters are sensitive to vertical light confinement. On the other hand, the 3-D method predicts an intrinsic Q of $7.7{\times}10^7$ and overall Q of $6.8{\times}10^7$, which is in good agreement with $6.3{\times}10^7$ and $5.5{\times}10^7$ obtained from the 2-D perturbation.  At large $d$ where the perturbation is negligible and solution consists of pure WGMs, the total Q between the 3-D ($7.6{\times}10^7$) and the 2-D ($6.8{\times}10^7$) reaches even closer agreement.

\section{Conclusion}
In conclusion, through numerical analysis with 2-D and 3-D perturbation methods, we confirmed that the polygon and star modes experimentally observed in~\cite{fang2020polygon,lin2022electro} are due to fiber-induced perturbation. We further verified our experimental observation that the polygon modes may reach high overall quality factors even with the fiber placed close to the disk surface, which in general will cause overcouple for conventional WGM.  Furthermore, although the 2-D perturbation method is a fast algorithm and provides reasonable accuracy in predicting field distribution and quality factors, it is less accurate in estimating parameters such as wedge angle and fiber-to-disk distance, which are highly sensitive to vertical light confinement. Under these situations, a more accurate 3-D perturbation approach is recommended.


\begin{thebibliography}{34}%
\makeatletter
\providecommand \@ifxundefined [1]{%
 \@ifx{#1\undefined}
}%
\providecommand \@ifnum [1]{%
 \ifnum #1\expandafter \@firstoftwo
 \else \expandafter \@secondoftwo
 \fi
}%
\providecommand \@ifx [1]{%
 \ifx #1\expandafter \@firstoftwo
 \else \expandafter \@secondoftwo
 \fi
}%
\providecommand \natexlab [1]{#1}%
\providecommand \enquote  [1]{``#1''}%
\providecommand \bibnamefont  [1]{#1}%
\providecommand \bibfnamefont [1]{#1}%
\providecommand \citenamefont [1]{#1}%
\providecommand \href@noop [0]{\@secondoftwo}%
\providecommand \href [0]{\begingroup \@sanitize@url \@href}%
\providecommand \@href[1]{\@@startlink{#1}\@@href}%
\providecommand \@@href[1]{\endgroup#1\@@endlink}%
\providecommand \@sanitize@url [0]{\catcode `\\12\catcode `\$12\catcode
  `\&12\catcode `\#12\catcode `\^12\catcode `\_12\catcode `\%12\relax}%
\providecommand \@@startlink[1]{}%
\providecommand \@@endlink[0]{}%
\providecommand \url  [0]{\begingroup\@sanitize@url \@url }%
\providecommand \@url [1]{\endgroup\@href {#1}{\urlprefix }}%
\providecommand \urlprefix  [0]{URL }%
\providecommand \Eprint [0]{\href }%
\providecommand \doibase [0]{https://doi.org/}%
\providecommand \selectlanguage [0]{\@gobble}%
\providecommand \bibinfo  [0]{\@secondoftwo}%
\providecommand \bibfield  [0]{\@secondoftwo}%
\providecommand \translation [1]{[#1]}%
\providecommand \BibitemOpen [0]{}%
\providecommand \bibitemStop [0]{}%
\providecommand \bibitemNoStop [0]{.\EOS\space}%
\providecommand \EOS [0]{\spacefactor3000\relax}%
\providecommand \BibitemShut  [1]{\csname bibitem#1\endcsname}%
\let\auto@bib@innerbib\@empty
\bibitem [{\citenamefont {Vahala}(2003)}]{vahala2003optical}%
  \BibitemOpen
  \bibfield  {author} {\bibinfo {author} {\bibfnamefont {K.~J.}\ \bibnamefont
  {Vahala}},\ }\bibfield  {title} {\bibinfo {title} {Optical microcavities},\
  }\href@noop {} {\bibfield  {journal} {\bibinfo  {journal} {Nature}\ }\textbf
  {\bibinfo {volume} {424}},\ \bibinfo {pages} {839} (\bibinfo {year}
  {2003})}\BibitemShut {NoStop}%
\bibitem [{\citenamefont {Lu}\ \emph {et~al.}(2011)\citenamefont {Lu},
  \citenamefont {Lee}, \citenamefont {Chen}, \citenamefont {Herchak},
  \citenamefont {Kim}, \citenamefont {Fraser}, \citenamefont {Flagan},\ and\
  \citenamefont {Vahala}}]{lu2011high}%
  \BibitemOpen
  \bibfield  {author} {\bibinfo {author} {\bibfnamefont {T.}~\bibnamefont
  {Lu}}, \bibinfo {author} {\bibfnamefont {H.}~\bibnamefont {Lee}}, \bibinfo
  {author} {\bibfnamefont {T.}~\bibnamefont {Chen}}, \bibinfo {author}
  {\bibfnamefont {S.}~\bibnamefont {Herchak}}, \bibinfo {author} {\bibfnamefont
  {J.-H.}\ \bibnamefont {Kim}}, \bibinfo {author} {\bibfnamefont {S.~E.}\
  \bibnamefont {Fraser}}, \bibinfo {author} {\bibfnamefont {R.~C.}\
  \bibnamefont {Flagan}},\ and\ \bibinfo {author} {\bibfnamefont
  {K.}~\bibnamefont {Vahala}},\ }\bibfield  {title} {\bibinfo {title} {High
  sensitivity nanoparticle detection using optical microcavities},\ }\href@noop
  {} {\bibfield  {journal} {\bibinfo  {journal} {Proceedings of the National
  Academy of Sciences}\ }\textbf {\bibinfo {volume} {108}},\ \bibinfo {pages}
  {5976} (\bibinfo {year} {2011})}\BibitemShut {NoStop}%
\bibitem [{\citenamefont {Baaske}\ \emph {et~al.}(2014)\citenamefont {Baaske},
  \citenamefont {Foreman},\ and\ \citenamefont {Vollmer}}]{baaske2014single}%
  \BibitemOpen
  \bibfield  {author} {\bibinfo {author} {\bibfnamefont {M.~D.}\ \bibnamefont
  {Baaske}}, \bibinfo {author} {\bibfnamefont {M.~R.}\ \bibnamefont
  {Foreman}},\ and\ \bibinfo {author} {\bibfnamefont {F.}~\bibnamefont
  {Vollmer}},\ }\bibfield  {title} {\bibinfo {title} {Single-molecule nucleic
  acid interactions monitored on a label-free microcavity biosensor platform},\
  }\href@noop {} {\bibfield  {journal} {\bibinfo  {journal} {Nature
  Nanotechnology}\ }\textbf {\bibinfo {volume} {9}},\ \bibinfo {pages} {933}
  (\bibinfo {year} {2014})}\BibitemShut {NoStop}%
\bibitem [{\citenamefont {Kippenberg}\ \emph {et~al.}(2011)\citenamefont
  {Kippenberg}, \citenamefont {Holzwarth},\ and\ \citenamefont
  {Diddams}}]{kippenberg2011microresonator}%
  \BibitemOpen
  \bibfield  {author} {\bibinfo {author} {\bibfnamefont {T.~J.}\ \bibnamefont
  {Kippenberg}}, \bibinfo {author} {\bibfnamefont {R.}~\bibnamefont
  {Holzwarth}},\ and\ \bibinfo {author} {\bibfnamefont {S.~A.}\ \bibnamefont
  {Diddams}},\ }\bibfield  {title} {\bibinfo {title} {Microresonator-based
  optical frequency combs},\ }\href@noop {} {\bibfield  {journal} {\bibinfo
  {journal} {Science}\ }\textbf {\bibinfo {volume} {332}},\ \bibinfo {pages}
  {555} (\bibinfo {year} {2011})}\BibitemShut {NoStop}%
\bibitem [{\citenamefont {Kippenberg}\ and\ \citenamefont
  {Vahala}(2008)}]{kippenberg2008cavity}%
  \BibitemOpen
  \bibfield  {author} {\bibinfo {author} {\bibfnamefont {T.~J.}\ \bibnamefont
  {Kippenberg}}\ and\ \bibinfo {author} {\bibfnamefont {K.~J.}\ \bibnamefont
  {Vahala}},\ }\bibfield  {title} {\bibinfo {title} {Cavity optomechanics:
  back-action at the mesoscale},\ }\href@noop {} {\bibfield  {journal}
  {\bibinfo  {journal} {Science}\ }\textbf {\bibinfo {volume} {321}},\ \bibinfo
  {pages} {1172} (\bibinfo {year} {2008})}\BibitemShut {NoStop}%
\bibitem [{\citenamefont {Honari}\ \emph {et~al.}(2021)\citenamefont {Honari},
  \citenamefont {Haque},\ and\ \citenamefont {Lu}}]{honari2021fabrication}%
  \BibitemOpen
  \bibfield  {author} {\bibinfo {author} {\bibfnamefont {S.}~\bibnamefont
  {Honari}}, \bibinfo {author} {\bibfnamefont {S.}~\bibnamefont {Haque}},\ and\
  \bibinfo {author} {\bibfnamefont {T.}~\bibnamefont {Lu}},\ }\bibfield
  {title} {\bibinfo {title} {Fabrication of ultra-high q silica microdisk using
  chemo-mechanical polishing},\ }\href@noop {} {\bibfield  {journal} {\bibinfo
  {journal} {Applied Physics Letters}\ }\textbf {\bibinfo {volume} {119}},\
  \bibinfo {pages} {031107} (\bibinfo {year} {2021})}\BibitemShut {NoStop}%
\bibitem [{\citenamefont {Yu}\ \emph {et~al.}(2016)\citenamefont {Yu},
  \citenamefont {Jiang}, \citenamefont {Lin},\ and\ \citenamefont
  {Lu}}]{yu2016cavity}%
  \BibitemOpen
  \bibfield  {author} {\bibinfo {author} {\bibfnamefont {W.}~\bibnamefont
  {Yu}}, \bibinfo {author} {\bibfnamefont {W.~C.}\ \bibnamefont {Jiang}},
  \bibinfo {author} {\bibfnamefont {Q.}~\bibnamefont {Lin}},\ and\ \bibinfo
  {author} {\bibfnamefont {T.}~\bibnamefont {Lu}},\ }\bibfield  {title}
  {\bibinfo {title} {Cavity optomechanical spring sensing of single
  molecules},\ }\href@noop {} {\bibfield  {journal} {\bibinfo  {journal}
  {Nature communications}\ }\textbf {\bibinfo {volume} {7}},\ \bibinfo {pages}
  {12311} (\bibinfo {year} {2016})}\BibitemShut {NoStop}%
\bibitem [{\citenamefont {Redding}\ \emph {et~al.}(2012)\citenamefont
  {Redding}, \citenamefont {Ge}, \citenamefont {Song}, \citenamefont {Wiersig},
  \citenamefont {Solomon},\ and\ \citenamefont {Cao}}]{redding2012local}%
  \BibitemOpen
  \bibfield  {author} {\bibinfo {author} {\bibfnamefont {B.}~\bibnamefont
  {Redding}}, \bibinfo {author} {\bibfnamefont {L.}~\bibnamefont {Ge}},
  \bibinfo {author} {\bibfnamefont {Q.}~\bibnamefont {Song}}, \bibinfo {author}
  {\bibfnamefont {J.}~\bibnamefont {Wiersig}}, \bibinfo {author} {\bibfnamefont
  {G.~S.}\ \bibnamefont {Solomon}},\ and\ \bibinfo {author} {\bibfnamefont
  {H.}~\bibnamefont {Cao}},\ }\bibfield  {title} {\bibinfo {title} {Local
  chirality of optical resonances in ultrasmall resonators},\ }\href@noop {}
  {\bibfield  {journal} {\bibinfo  {journal} {Physical Review Letters}\
  }\textbf {\bibinfo {volume} {108}},\ \bibinfo {pages} {253902} (\bibinfo
  {year} {2012})}\BibitemShut {NoStop}%
\bibitem [{\citenamefont {Fang}\ \emph {et~al.}(2007)\citenamefont {Fang},
  \citenamefont {Cao},\ and\ \citenamefont {Solomon}}]{fang2007control}%
  \BibitemOpen
  \bibfield  {author} {\bibinfo {author} {\bibfnamefont {W.}~\bibnamefont
  {Fang}}, \bibinfo {author} {\bibfnamefont {H.}~\bibnamefont {Cao}},\ and\
  \bibinfo {author} {\bibfnamefont {G.~S.}\ \bibnamefont {Solomon}},\
  }\bibfield  {title} {\bibinfo {title} {Control of lasing in fully chaotic
  open microcavities by tailoring the shape factor},\ }\href@noop {} {\bibfield
   {journal} {\bibinfo  {journal} {Applied Physics Letters}\ }\textbf {\bibinfo
  {volume} {90}},\ \bibinfo {pages} {081108} (\bibinfo {year}
  {2007})}\BibitemShut {NoStop}%
\bibitem [{\citenamefont {Unterhinninghofen}\ \emph {et~al.}(2008)\citenamefont
  {Unterhinninghofen}, \citenamefont {Wiersig},\ and\ \citenamefont
  {Hentschel}}]{unterhinninghofen2008goos}%
  \BibitemOpen
  \bibfield  {author} {\bibinfo {author} {\bibfnamefont {J.}~\bibnamefont
  {Unterhinninghofen}}, \bibinfo {author} {\bibfnamefont {J.}~\bibnamefont
  {Wiersig}},\ and\ \bibinfo {author} {\bibfnamefont {M.}~\bibnamefont
  {Hentschel}},\ }\bibfield  {title} {\bibinfo {title} {Goos-h{\"a}nchen shift
  and localization of optical modes in deformed microcavities},\ }\href@noop {}
  {\bibfield  {journal} {\bibinfo  {journal} {Physical Review E}\ }\textbf
  {\bibinfo {volume} {78}},\ \bibinfo {pages} {016201} (\bibinfo {year}
  {2008})}\BibitemShut {NoStop}%
\bibitem [{\citenamefont {Harayama}\ \emph {et~al.}(2005)\citenamefont
  {Harayama}, \citenamefont {Sunada},\ and\ \citenamefont
  {Ikeda}}]{harayama2005theory}%
  \BibitemOpen
  \bibfield  {author} {\bibinfo {author} {\bibfnamefont {T.}~\bibnamefont
  {Harayama}}, \bibinfo {author} {\bibfnamefont {S.}~\bibnamefont {Sunada}},\
  and\ \bibinfo {author} {\bibfnamefont {K.~S.}\ \bibnamefont {Ikeda}},\
  }\bibfield  {title} {\bibinfo {title} {Theory of two-dimensional microcavity
  lasers},\ }\href@noop {} {\bibfield  {journal} {\bibinfo  {journal} {Physical
  Review A}\ }\textbf {\bibinfo {volume} {72}},\ \bibinfo {pages} {013803}
  (\bibinfo {year} {2005})}\BibitemShut {NoStop}%
\bibitem [{\citenamefont {Wiersig}\ and\ \citenamefont
  {Hentschel}(2008)}]{wiersig2008combining}%
  \BibitemOpen
  \bibfield  {author} {\bibinfo {author} {\bibfnamefont {J.}~\bibnamefont
  {Wiersig}}\ and\ \bibinfo {author} {\bibfnamefont {M.}~\bibnamefont
  {Hentschel}},\ }\bibfield  {title} {\bibinfo {title} {Combining directional
  light output and ultralow loss in deformed microdisks},\ }\href@noop {}
  {\bibfield  {journal} {\bibinfo  {journal} {Physical Review Letters}\
  }\textbf {\bibinfo {volume} {100}},\ \bibinfo {pages} {033901} (\bibinfo
  {year} {2008})}\BibitemShut {NoStop}%
\bibitem [{\citenamefont {Rex}\ \emph {et~al.}(2002)\citenamefont {Rex},
  \citenamefont {Tureci}, \citenamefont {Schwefel}, \citenamefont {Chang},\
  and\ \citenamefont {Stone}}]{rex2002fresnel}%
  \BibitemOpen
  \bibfield  {author} {\bibinfo {author} {\bibfnamefont {N.}~\bibnamefont
  {Rex}}, \bibinfo {author} {\bibfnamefont {H.~E.}\ \bibnamefont {Tureci}},
  \bibinfo {author} {\bibfnamefont {H.}~\bibnamefont {Schwefel}}, \bibinfo
  {author} {\bibfnamefont {R.}~\bibnamefont {Chang}},\ and\ \bibinfo {author}
  {\bibfnamefont {A.~D.}\ \bibnamefont {Stone}},\ }\bibfield  {title} {\bibinfo
  {title} {Fresnel filtering in lasing emission from scarred modes of
  wave-chaotic optical resonators},\ }\href@noop {} {\bibfield  {journal}
  {\bibinfo  {journal} {Physical Review Letters}\ }\textbf {\bibinfo {volume}
  {88}},\ \bibinfo {pages} {094102} (\bibinfo {year} {2002})}\BibitemShut
  {NoStop}%
\bibitem [{\citenamefont {Fang}\ \emph {et~al.}(2020)\citenamefont {Fang},
  \citenamefont {Haque}, \citenamefont {Farajollahi}, \citenamefont {Luo},
  \citenamefont {Lin}, \citenamefont {Wu}, \citenamefont {Zhang}, \citenamefont
  {Wang}, \citenamefont {Wang}, \citenamefont {Cheng} \emph
  {et~al.}}]{fang2020polygon}%
  \BibitemOpen
  \bibfield  {author} {\bibinfo {author} {\bibfnamefont {Z.}~\bibnamefont
  {Fang}}, \bibinfo {author} {\bibfnamefont {S.}~\bibnamefont {Haque}},
  \bibinfo {author} {\bibfnamefont {S.}~\bibnamefont {Farajollahi}}, \bibinfo
  {author} {\bibfnamefont {H.}~\bibnamefont {Luo}}, \bibinfo {author}
  {\bibfnamefont {J.}~\bibnamefont {Lin}}, \bibinfo {author} {\bibfnamefont
  {R.}~\bibnamefont {Wu}}, \bibinfo {author} {\bibfnamefont {J.}~\bibnamefont
  {Zhang}}, \bibinfo {author} {\bibfnamefont {Z.}~\bibnamefont {Wang}},
  \bibinfo {author} {\bibfnamefont {M.}~\bibnamefont {Wang}}, \bibinfo {author}
  {\bibfnamefont {Y.}~\bibnamefont {Cheng}}, \emph {et~al.},\ }\bibfield
  {title} {\bibinfo {title} {Polygon coherent modes in a weakly perturbed
  whispering gallery microresonator for efficient second harmonic,
  optomechanical, and frequency comb generations},\ }\href@noop {} {\bibfield
  {journal} {\bibinfo  {journal} {Physical Review Letters}\ }\textbf {\bibinfo
  {volume} {125}},\ \bibinfo {pages} {173901} (\bibinfo {year}
  {2020})}\BibitemShut {NoStop}%
\bibitem [{\citenamefont {Lin}\ \emph {et~al.}(2022)\citenamefont {Lin},
  \citenamefont {Farajollahi}, \citenamefont {Fang}, \citenamefont {Yao},
  \citenamefont {Gao}, \citenamefont {Guan}, \citenamefont {Deng},
  \citenamefont {Lu}, \citenamefont {Wang}, \citenamefont {Zhang} \emph
  {et~al.}}]{lin2022electro}%
  \BibitemOpen
  \bibfield  {author} {\bibinfo {author} {\bibfnamefont {J.}~\bibnamefont
  {Lin}}, \bibinfo {author} {\bibfnamefont {S.}~\bibnamefont {Farajollahi}},
  \bibinfo {author} {\bibfnamefont {Z.}~\bibnamefont {Fang}}, \bibinfo {author}
  {\bibfnamefont {N.}~\bibnamefont {Yao}}, \bibinfo {author} {\bibfnamefont
  {R.}~\bibnamefont {Gao}}, \bibinfo {author} {\bibfnamefont {J.}~\bibnamefont
  {Guan}}, \bibinfo {author} {\bibfnamefont {L.}~\bibnamefont {Deng}}, \bibinfo
  {author} {\bibfnamefont {T.}~\bibnamefont {Lu}}, \bibinfo {author}
  {\bibfnamefont {M.}~\bibnamefont {Wang}}, \bibinfo {author} {\bibfnamefont
  {H.}~\bibnamefont {Zhang}}, \emph {et~al.},\ }\bibfield  {title} {\bibinfo
  {title} {Electro-optic tuning of a single-frequency ultranarrow linewidth
  microdisk laser},\ }\href@noop {} {\bibfield  {journal} {\bibinfo  {journal}
  {Advanced Photonics}\ }\textbf {\bibinfo {volume} {4}},\ \bibinfo {pages}
  {036001} (\bibinfo {year} {2022})}\BibitemShut {NoStop}%
\bibitem [{\citenamefont {Lee}\ \emph {et~al.}(2004)\citenamefont {Lee},
  \citenamefont {Rim}, \citenamefont {Ryu}, \citenamefont {Kwon}, \citenamefont
  {Choi},\ and\ \citenamefont {Kim}}]{lee2004quasiscarred}%
  \BibitemOpen
  \bibfield  {author} {\bibinfo {author} {\bibfnamefont {S.-Y.}\ \bibnamefont
  {Lee}}, \bibinfo {author} {\bibfnamefont {S.}~\bibnamefont {Rim}}, \bibinfo
  {author} {\bibfnamefont {J.-W.}\ \bibnamefont {Ryu}}, \bibinfo {author}
  {\bibfnamefont {T.-Y.}\ \bibnamefont {Kwon}}, \bibinfo {author}
  {\bibfnamefont {M.}~\bibnamefont {Choi}},\ and\ \bibinfo {author}
  {\bibfnamefont {C.-M.}\ \bibnamefont {Kim}},\ }\bibfield  {title} {\bibinfo
  {title} {Quasiscarred resonances in a spiral-shaped microcavity},\
  }\href@noop {} {\bibfield  {journal} {\bibinfo  {journal} {Physical Review
  Letters}\ }\textbf {\bibinfo {volume} {93}},\ \bibinfo {pages} {164102}
  (\bibinfo {year} {2004})}\BibitemShut {NoStop}%
\bibitem [{\citenamefont {Wiersig}(2002)}]{wiersig2002boundary}%
  \BibitemOpen
  \bibfield  {author} {\bibinfo {author} {\bibfnamefont {J.}~\bibnamefont
  {Wiersig}},\ }\bibfield  {title} {\bibinfo {title} {Boundary element method
  for resonances in dielectric microcavities},\ }\href@noop {} {\bibfield
  {journal} {\bibinfo  {journal} {Journal of Optics A: Pure and Applied
  Optics}\ }\textbf {\bibinfo {volume} {5}},\ \bibinfo {pages} {53} (\bibinfo
  {year} {2002})}\BibitemShut {NoStop}%
\bibitem [{\citenamefont {Zou}\ \emph {et~al.}(2009)\citenamefont {Zou},
  \citenamefont {Yang}, \citenamefont {Xiao}, \citenamefont {Dong},
  \citenamefont {Han},\ and\ \citenamefont {Guo}}]{zou2009accurately}%
  \BibitemOpen
  \bibfield  {author} {\bibinfo {author} {\bibfnamefont {C.-L.}\ \bibnamefont
  {Zou}}, \bibinfo {author} {\bibfnamefont {Y.}~\bibnamefont {Yang}}, \bibinfo
  {author} {\bibfnamefont {Y.-F.}\ \bibnamefont {Xiao}}, \bibinfo {author}
  {\bibfnamefont {C.-H.}\ \bibnamefont {Dong}}, \bibinfo {author}
  {\bibfnamefont {Z.-F.}\ \bibnamefont {Han}},\ and\ \bibinfo {author}
  {\bibfnamefont {G.-C.}\ \bibnamefont {Guo}},\ }\bibfield  {title} {\bibinfo
  {title} {Accurately calculating high quality factor of whispering-gallery
  modes with boundary element method},\ }\href@noop {} {\bibfield  {journal}
  {\bibinfo  {journal} {JOSA B}\ }\textbf {\bibinfo {volume} {26}},\ \bibinfo
  {pages} {2050} (\bibinfo {year} {2009})}\BibitemShut {NoStop}%
\bibitem [{\citenamefont {Zou}\ \emph {et~al.}(2011)\citenamefont {Zou},
  \citenamefont {Schwefel}, \citenamefont {Sun}, \citenamefont {Han},\ and\
  \citenamefont {Guo}}]{zou2011quick}%
  \BibitemOpen
  \bibfield  {author} {\bibinfo {author} {\bibfnamefont {C.-L.}\ \bibnamefont
  {Zou}}, \bibinfo {author} {\bibfnamefont {H.~G.}\ \bibnamefont {Schwefel}},
  \bibinfo {author} {\bibfnamefont {F.-W.}\ \bibnamefont {Sun}}, \bibinfo
  {author} {\bibfnamefont {Z.-F.}\ \bibnamefont {Han}},\ and\ \bibinfo {author}
  {\bibfnamefont {G.-C.}\ \bibnamefont {Guo}},\ }\bibfield  {title} {\bibinfo
  {title} {Quick root searching method for resonances of dielectric optical
  microcavities with the boundary element method},\ }\href@noop {} {\bibfield
  {journal} {\bibinfo  {journal} {Optics Express}\ }\textbf {\bibinfo {volume}
  {19}},\ \bibinfo {pages} {15669} (\bibinfo {year} {2011})}\BibitemShut
  {NoStop}%
\bibitem [{\citenamefont {Teraoka}\ \emph {et~al.}(2003)\citenamefont
  {Teraoka}, \citenamefont {Arnold},\ and\ \citenamefont
  {Vollmer}}]{teraoka2003perturbation}%
  \BibitemOpen
  \bibfield  {author} {\bibinfo {author} {\bibfnamefont {I.}~\bibnamefont
  {Teraoka}}, \bibinfo {author} {\bibfnamefont {S.}~\bibnamefont {Arnold}},\
  and\ \bibinfo {author} {\bibfnamefont {F.}~\bibnamefont {Vollmer}},\
  }\bibfield  {title} {\bibinfo {title} {Perturbation approach to resonance
  shifts of whispering-gallery modes in a dielectric microsphere as a probe of
  a surrounding medium},\ }\href@noop {} {\bibfield  {journal} {\bibinfo
  {journal} {JOSA B}\ }\textbf {\bibinfo {volume} {20}},\ \bibinfo {pages}
  {1937} (\bibinfo {year} {2003})}\BibitemShut {NoStop}%
\bibitem [{\citenamefont {Arnold}\ \emph {et~al.}(2003)\citenamefont {Arnold},
  \citenamefont {Khoshsima}, \citenamefont {Teraoka}, \citenamefont {Holler},\
  and\ \citenamefont {Vollmer}}]{arnold2003shift}%
  \BibitemOpen
  \bibfield  {author} {\bibinfo {author} {\bibfnamefont {S.}~\bibnamefont
  {Arnold}}, \bibinfo {author} {\bibfnamefont {M.}~\bibnamefont {Khoshsima}},
  \bibinfo {author} {\bibfnamefont {I.}~\bibnamefont {Teraoka}}, \bibinfo
  {author} {\bibfnamefont {S.}~\bibnamefont {Holler}},\ and\ \bibinfo {author}
  {\bibfnamefont {F.}~\bibnamefont {Vollmer}},\ }\bibfield  {title} {\bibinfo
  {title} {Shift of whispering-gallery modes in microspheres by protein
  adsorption},\ }\href@noop {} {\bibfield  {journal} {\bibinfo  {journal}
  {Optics Letters}\ }\textbf {\bibinfo {volume} {28}},\ \bibinfo {pages} {272}
  (\bibinfo {year} {2003})}\BibitemShut {NoStop}%
\bibitem [{\citenamefont {Teraoka}\ and\ \citenamefont
  {Arnold}(2006)}]{teraoka2006theory}%
  \BibitemOpen
  \bibfield  {author} {\bibinfo {author} {\bibfnamefont {I.}~\bibnamefont
  {Teraoka}}\ and\ \bibinfo {author} {\bibfnamefont {S.}~\bibnamefont
  {Arnold}},\ }\bibfield  {title} {\bibinfo {title} {Theory of resonance shifts
  in te and tm whispering gallery modes by nonradial perturbations for sensing
  applications},\ }\href@noop {} {\bibfield  {journal} {\bibinfo  {journal}
  {JOSA B}\ }\textbf {\bibinfo {volume} {23}},\ \bibinfo {pages} {1381}
  (\bibinfo {year} {2006})}\BibitemShut {NoStop}%
\bibitem [{\citenamefont {Foreman}\ and\ \citenamefont
  {Vollmer}(2013)}]{foreman2013theory}%
  \BibitemOpen
  \bibfield  {author} {\bibinfo {author} {\bibfnamefont {M.~R.}\ \bibnamefont
  {Foreman}}\ and\ \bibinfo {author} {\bibfnamefont {F.}~\bibnamefont
  {Vollmer}},\ }\bibfield  {title} {\bibinfo {title} {Theory of resonance
  shifts of whispering gallery modes by arbitrary plasmonic nanoparticles},\
  }\href@noop {} {\bibfield  {journal} {\bibinfo  {journal} {New Journal of
  Physics}\ }\textbf {\bibinfo {volume} {15}},\ \bibinfo {pages} {083006}
  (\bibinfo {year} {2013})}\BibitemShut {NoStop}%
\bibitem [{\citenamefont {Swaim}\ \emph {et~al.}(2011)\citenamefont {Swaim},
  \citenamefont {Knittel},\ and\ \citenamefont {Bowen}}]{swaim2011detection}%
  \BibitemOpen
  \bibfield  {author} {\bibinfo {author} {\bibfnamefont {J.~D.}\ \bibnamefont
  {Swaim}}, \bibinfo {author} {\bibfnamefont {J.}~\bibnamefont {Knittel}},\
  and\ \bibinfo {author} {\bibfnamefont {W.~P.}\ \bibnamefont {Bowen}},\
  }\bibfield  {title} {\bibinfo {title} {Detection limits in whispering gallery
  biosensors with plasmonic enhancement},\ }\href@noop {} {\bibfield  {journal}
  {\bibinfo  {journal} {Applied Physics Letters}\ }\textbf {\bibinfo {volume}
  {99}},\ \bibinfo {pages} {243109} (\bibinfo {year} {2011})}\BibitemShut
  {NoStop}%
\bibitem [{\citenamefont {Lai}\ \emph {et~al.}(1990)\citenamefont {Lai},
  \citenamefont {Leung}, \citenamefont {Young}, \citenamefont {Barber},\ and\
  \citenamefont {Hill}}]{lai1990time}%
  \BibitemOpen
  \bibfield  {author} {\bibinfo {author} {\bibfnamefont {H.}~\bibnamefont
  {Lai}}, \bibinfo {author} {\bibfnamefont {P.}~\bibnamefont {Leung}}, \bibinfo
  {author} {\bibfnamefont {K.}~\bibnamefont {Young}}, \bibinfo {author}
  {\bibfnamefont {P.}~\bibnamefont {Barber}},\ and\ \bibinfo {author}
  {\bibfnamefont {S.}~\bibnamefont {Hill}},\ }\bibfield  {title} {\bibinfo
  {title} {Time-independent perturbation for leaking electromagnetic modes in
  open systems with application to resonances in microdroplets},\ }\href@noop
  {} {\bibfield  {journal} {\bibinfo  {journal} {Physical Review A}\ }\textbf
  {\bibinfo {volume} {41}},\ \bibinfo {pages} {5187} (\bibinfo {year}
  {1990})}\BibitemShut {NoStop}%
\bibitem [{\citenamefont {Lee}\ \emph {et~al.}(2008)\citenamefont {Lee},
  \citenamefont {Rim}, \citenamefont {Cho},\ and\ \citenamefont
  {Kim}}]{lee2008resonances}%
  \BibitemOpen
  \bibfield  {author} {\bibinfo {author} {\bibfnamefont {J.}~\bibnamefont
  {Lee}}, \bibinfo {author} {\bibfnamefont {S.}~\bibnamefont {Rim}}, \bibinfo
  {author} {\bibfnamefont {J.}~\bibnamefont {Cho}},\ and\ \bibinfo {author}
  {\bibfnamefont {C.-M.}\ \bibnamefont {Kim}},\ }\bibfield  {title} {\bibinfo
  {title} {Resonances near the classical separatrix of a weakly deformed
  circular microcavity},\ }\href@noop {} {\bibfield  {journal} {\bibinfo
  {journal} {Physical Review Letters}\ }\textbf {\bibinfo {volume} {101}},\
  \bibinfo {pages} {064101} (\bibinfo {year} {2008})}\BibitemShut {NoStop}%
\bibitem [{\citenamefont {T{\"u}reci}\ \emph {et~al.}(2005)\citenamefont
  {T{\"u}reci}, \citenamefont {Schwefel}, \citenamefont {Jacquod},\ and\
  \citenamefont {Stone}}]{tureci2005modes}%
  \BibitemOpen
  \bibfield  {author} {\bibinfo {author} {\bibfnamefont {H.}~\bibnamefont
  {T{\"u}reci}}, \bibinfo {author} {\bibfnamefont {H.}~\bibnamefont
  {Schwefel}}, \bibinfo {author} {\bibfnamefont {P.}~\bibnamefont {Jacquod}},\
  and\ \bibinfo {author} {\bibfnamefont {A.~D.}\ \bibnamefont {Stone}},\
  }\bibfield  {title} {\bibinfo {title} {Modes of wave-chaotic dielectric
  resonators},\ }in\ \href@noop {} {\emph {\bibinfo {booktitle} {Progress in
  Optics}}}\ (\bibinfo  {publisher} {Elsevier},\ \bibinfo {year} {2005})\ pp.\
  \bibinfo {pages} {75--137}\BibitemShut {NoStop}%
\bibitem [{\citenamefont {Korneev}(2016)}]{korneev2016perturbation}%
  \BibitemOpen
  \bibfield  {author} {\bibinfo {author} {\bibfnamefont {N.}~\bibnamefont
  {Korneev}},\ }\bibfield  {title} {\bibinfo {title} {Perturbation
  approximation for higher modes in nearly regular two-dimensional cavities},\
  }\href@noop {} {\bibfield  {journal} {\bibinfo  {journal} {Cogent Physics}\
  }\textbf {\bibinfo {volume} {3}},\ \bibinfo {pages} {1262725} (\bibinfo
  {year} {2016})}\BibitemShut {NoStop}%
\bibitem [{\citenamefont {Gorodetsky}\ and\ \citenamefont
  {Ilchenko}(1999)}]{gorodetsky1999optical}%
  \BibitemOpen
  \bibfield  {author} {\bibinfo {author} {\bibfnamefont {M.~L.}\ \bibnamefont
  {Gorodetsky}}\ and\ \bibinfo {author} {\bibfnamefont {V.~S.}\ \bibnamefont
  {Ilchenko}},\ }\bibfield  {title} {\bibinfo {title} {Optical microsphere
  resonators: optimal coupling to high-{Q} whispering-gallery modes},\
  }\href@noop {} {\bibfield  {journal} {\bibinfo  {journal} {JOSA B}\ }\textbf
  {\bibinfo {volume} {16}},\ \bibinfo {pages} {147} (\bibinfo {year}
  {1999})}\BibitemShut {NoStop}%
\bibitem [{\citenamefont {Du}\ \emph {et~al.}(2013)\citenamefont {Du},
  \citenamefont {Vincent},\ and\ \citenamefont {Lu}}]{du2013full}%
  \BibitemOpen
  \bibfield  {author} {\bibinfo {author} {\bibfnamefont {X.}~\bibnamefont
  {Du}}, \bibinfo {author} {\bibfnamefont {S.}~\bibnamefont {Vincent}},\ and\
  \bibinfo {author} {\bibfnamefont {T.}~\bibnamefont {Lu}},\ }\bibfield
  {title} {\bibinfo {title} {Full-vectorial whispering-gallery-mode cavity
  analysis},\ }\href@noop {} {\bibfield  {journal} {\bibinfo  {journal} {Optics
  Express}\ }\textbf {\bibinfo {volume} {21}},\ \bibinfo {pages} {22012}
  (\bibinfo {year} {2013})}\BibitemShut {NoStop}%
\bibitem [{\citenamefont {Du}\ \emph {et~al.}(2014)\citenamefont {Du},
  \citenamefont {Vincent}, \citenamefont {Faucher}, \citenamefont {Picard},\
  and\ \citenamefont {Lu}}]{du2014generalized}%
  \BibitemOpen
  \bibfield  {author} {\bibinfo {author} {\bibfnamefont {X.}~\bibnamefont
  {Du}}, \bibinfo {author} {\bibfnamefont {S.}~\bibnamefont {Vincent}},
  \bibinfo {author} {\bibfnamefont {M.}~\bibnamefont {Faucher}}, \bibinfo
  {author} {\bibfnamefont {M.-J.}\ \bibnamefont {Picard}},\ and\ \bibinfo
  {author} {\bibfnamefont {T.}~\bibnamefont {Lu}},\ }\bibfield  {title}
  {\bibinfo {title} {Generalized full-vector multi-mode matching analysis of
  whispering gallery microcavities},\ }\href@noop {} {\bibfield  {journal}
  {\bibinfo  {journal} {Optics Express}\ }\textbf {\bibinfo {volume} {22}},\
  \bibinfo {pages} {13507} (\bibinfo {year} {2014})}\BibitemShut {NoStop}%
\bibitem [{\citenamefont {Rowland}\ and\ \citenamefont
  {Love}(1993)}]{rowland1993evanescent}%
  \BibitemOpen
  \bibfield  {author} {\bibinfo {author} {\bibfnamefont {D.}~\bibnamefont
  {Rowland}}\ and\ \bibinfo {author} {\bibfnamefont {J.}~\bibnamefont {Love}},\
  }\bibfield  {title} {\bibinfo {title} {Evanescent wave coupling of whispering
  gallery modes of a dielectric cylinder},\ }\href@noop {} {\bibfield
  {journal} {\bibinfo  {journal} {IEE Proceedings J (Optoelectronics)}\
  }\textbf {\bibinfo {volume} {140}},\ \bibinfo {pages} {177} (\bibinfo {year}
  {1993})}\BibitemShut {NoStop}%
\bibitem [{\citenamefont {Snyder}\ and\ \citenamefont
  {Love}(2012)}]{snyder2012optical}%
  \BibitemOpen
  \bibfield  {author} {\bibinfo {author} {\bibfnamefont {A.~W.}\ \bibnamefont
  {Snyder}}\ and\ \bibinfo {author} {\bibfnamefont {J.}~\bibnamefont {Love}},\
  }\href@noop {} {\emph {\bibinfo {title} {Optical waveguide theory}}}\
  (\bibinfo  {publisher} {Springer Science \& Business Media},\ \bibinfo {year}
  {2012})\BibitemShut {NoStop}%
\end{thebibliography}
\providecommand{\noopsort}[1]{}\providecommand{\singleletter}[1]{#1}%

\end{document}